\documentclass[preprint,review,12pt]{elsarticle}

\usepackage{amssymb}
\usepackage[utf8]{inputenc}
\usepackage[T1]{fontenc}
\usepackage{lmodern}
\usepackage{graphicx}
\usepackage[figurename=Fig.,labelfont=bf,labelsep=period]{caption}
\usepackage{subcaption}
\usepackage{amsmath}
\usepackage{upgreek}
\usepackage{newtxtext,newtxmath}
\usepackage[colorlinks=true,citecolor=red,linkcolor=black]{hyperref}
\usepackage{enumitem}

\newcommand{\bc}{\begin{center}}
	\newcommand{\ec}{\end{center}}
\newcommand{\bfr}{\begin{flushright}}
	\newcommand{\efr}{\end{flushright}}

\newcommand{\be}{\begin{enumerate}}
	\newcommand{\ee}{\end{enumerate}}
\newcommand{\bi}{\begin{itemize}}
	\newcommand{\ei}{\end{itemize}}
\newcommand{\bd}{\begin{description}}
	\newcommand{\ed}{\end{description}}
\newcommand{\beq}{\begin{equation}}
\newcommand{\eeq}{\end{equation}}
\newcommand{\bea}{\begin{eqnarray}}
\newcommand{\eea}{\end{eqnarray}}

\newcommand{\bfi}{\begin{figure}}
	\newcommand{\efi}{\end{figure}}
\newcommand{\bay}{\begin{array}{l}}
	\newcommand{\eay}{\end{array}}

\def\mb#1{\mbox {\boldmath {$#1$}}} 

\usepackage{tikz}

\journal{Engineering Fracture Mechanics}

\begin{document}

\begin{frontmatter}



\title{Homogenization Coarse Graining (HCG) of the Lattice Discrete Particle Model (LDPM) for the Analysis of Reinforced Concrete Structures}


\author[label1]{Erol Lale}
\author[label2]{Roozbeh Rezakhani}
\author[label3]{Mohammed Alnaggar}
\author[label2]{Gianluca Cusatis}

\address[label1]{Department of Civil Engineering, Istanbul Technical University, Istanbul, Turkey.}
\address[label2]{Department of Civil and Environmental Engineering, Northwestern University, Evanston (IL), USA.}
\address[label3]{Department of Civil and Environmental Engineering, Rensselaer Polytechnic Institute, Troy (NY), USA.}

\begin{abstract}
In this study, a coarse-graining framework for discrete models is formulated on the basis of multiscale homogenization. The discrete model considered in this paper is the Lattice Discrete Particle Model (LDPM), which simulates concrete at the level of coarse aggregate pieces. In LDPM, the size of the aggregate particles follows the actual particle size distribution that is used in experiment to produce concrete specimens. Consequently, modeling large structural systems entirely with LDPM leads to a tremendous number of degrees of freedom and is not feasible with the currently available computational resources. To overcome this limitation, this paper proposes the formulation of a  coarse-grained model obtained by (1) increasing the actual size of the particles in the fine-scale model by a specific coarsening factor and (2) calibrating the parameters of the coarse grained model by best fitting the macroscopic, average response of the coarse grained model to the corresponding fine scale one for different loading conditions. A Representative Volume Element (RVE) of LDPM is employed to obtain the macroscopic response of the fine scale and coarse grained models through a homogenization procedure. Accuracy and efficiency of the developed coarse graining method is verified by comparing the response of fine  scale and coarse grained simulations of several reinforced concrete structural systems in terms of both accuracy of the results and computational cost.

\end{abstract}


\end{frontmatter}


\section{Introduction}

Cementitious composites materials, such as concrete, are widely used in engineering applications. These materials are heterogeneous, are characterized by quasi-brittle mechanical behavior, and their mechanical response is strongly influenced by various phenomena  such as crack initiation and propagation, interaction and coalescence of distributed micro-cracks into a localized macro crack, existence of confining pressure, and crack bridging effect of fibers. These phenomena occur at different spatial scales ranging from atomistic scale ($\sim10^{-10}$ m) to the structural scale ($\sim10^1$ m). Several micro-, meso-, and macro-scale constitutive models have been developed to simulate this complex behavior at various scales. Most of the developed models are based on continuum mechanics which neglect the complex internal structure of the material, They are well suited to capture the global response of a structure, when inelastic behavior is distributed over a large volume of material. However, these models become inaccurate for complex loading conditions in which macroscopic mechanical behavior is heavily influenced by material heterogeneity and damage localization. 


Mini-scale models were proposed by several authors including Wittmann and coworkers \cite{roelfstra1985beton} in 2D and Carol and coworkers \cite{caballero2006meso, caballero2006new, caballero20063d} in 3D. They used the finite element method to discretize coarse aggregate pieces, mortar matrix, and aggregate-matrix interface. As an alternative to the use of continuum approaches, Schlangen modeled concrete through a discrete system of beams (lattice elements) \cite{schlangen1992simple}. In his approach, lattice meshes were used to create the internal structure of concrete, in which different material properties were assigned to the lattice elements corresponding to the various components such as matrix, aggregate, and interface. Bolander and co-workers \cite{bolander1998fracture, yip2006irregular}  formulated a discrete mini-scale model based on the interaction between rigid polyhedral particles constructed through the Voronoi tessellation of the domain. Similar approach was employed by Nagai et al. \cite{nagai2004mesoscopic} to simulate mortar and concrete in a 2D setting. Mini-scale models provide realistic simulations of concrete cracking, coalescence of several distributed cracks into a localized one, and fracture propagation.  However, these models tend to be computationally intensive even to simulate laboratory test, and for 3D modeling which is necessary to capture correctly compressive failure and confinement effects.


The effect of the material internal structure on the macroscopic behavior can be analyzed efficiently by meso-scale model employing particle models in which only coarse aggregate pieces are simulated, and each particle corresponds to a single aggregate. This approach was applied successfully to geomaterials \cite{cundall1979discrete, plesha1983modeling} as well as concrete \cite{zubelewicz1987interface, cusatis2003confinement,Donze-discrete}. Meso-scale models reduce considerably the size of the numerical problem, while they can capture the fundamental aspects of material heterogeneity along with damage localization and fracture processes even in the case of three-dimensional complex fracture phenomena.

Building upon earlier developments \citep{cusatis2003confinement,cusatis2003confinementII}, Cusatis and coworkers developed an efficient mesoscale model for the simulation of concrete: the so-called Lattice Discrete Particle Model (LDPM) \cite{cusatis-ldpm-1,cusatis-ldpm-2}. LDPM simulates concrete internal structure by modeling coarse aggregate pieces and approximating their interaction through the interaction of rigid polyhedral cells. LDPM successfully simulate concrete mechanical behavior by employing meso-scale constitutive relationships in which three major failue mechanisms are taken into account: fracture and cohesion in tension; compaction and pore collapse under compression; and frictional behavior in shear. LDPM provides a computationally efficient framework, which is able to model most aspects of concrete behavior such as uniaxial, biaxial, and triaxial responses. 


Although the meso-scale modeling of various experiments on laboratory size concrete specimen has been performed successfully by particle models \cite{cusatis-ldpm-2}, numerical simulation of real size engineering structures using meso-scale models is impractical even with the use of parallel computing techniques. For instance, a concrete cylindrical specimen of 150 mm diameter and 300 mm height with maximum aggregate size of 10 mm simulated with LDPM includes approximately 8,500 particles. This yields to 51,000 degrees of freedom, given that each node has 6 degrees of freedom. This clearly shows that the simulation of large concrete structures using LDPM is demanding computationally, as it requires solving a computational system characterized by billions of degrees of freedom \cite{Alnaggar2012automatic}. Therefore, during the past few decades, researchers have developed multiscale computational methods, by which numerical simulation of large engineering problems is feasible within a reasonale amount of computational cost. 


Among various multiscale models, computational multiscale homogenization methods have been extensively studied and sucessfully employed for the simulation of different heterogeneous materials. Multiscale homogenization method is a hierarchical approach, in which at least two scales of problem are considered simultaneously. At the lower scale, the heterogeneous structure of the material is simulated explicitly in a certain volume, the so-called Representative Volume Element (RVE), which carries a complete information of the internal structure \cite{gitman2007representative, kouznetsova2004size, Bostanabad-1,Bessa-1}. At the macro-scale, the material is considered to be homogeneous, and during the analysis information flows between the two scales \cite{smit1998prediction, kouznetsova2004size, miehe1999computational}. In this approach, the macroscopic material domain is discretized by finite elements, and a single RVE is assigned to each Gauss point of the macroscopic finite elements. At each computational step, strains at macro-level are imposed as essential boundary conditions to the corresponding RVE, and the solution of the RVE boundary value problem is then averaged for the calculation of the associated macroscopic stress tensor. The Asymptotic Expansion Homogenization (AEH) is a similar but mathematically more rigorous and it exploits the asymptotic expansion of the displacement field considering a length scale parameter representing the ratio of the material heterogeneity length scale to the macroscopic one \citep{Hasani-1,Hasani-2}. Fish et al. \cite{fish2007generalized} presented a generalized formulation for this approach and introduced the Generalized Mathematical Homogenization (GMH) for the homogenization of atomistic systems. Based on GMH, Rezakhani and Cusatis \cite{rezakhani2016asymptotic,rezakhani2017adaptive} derived a homogenization scheme for discrete models featuring both translational and rotational degrees of freedom.

Coarse Graining (CG) methods are another class of multiscale methods for reducing the computational cost of discrete Fine Scale (FS) models. The method is based on converting a model with large number of degrees of freedom into a model with a reduced number of degrees of freedom but with the same mathematical and computational structure. The computational gain of CG is two fold: (1) the decrease of the number of degrees of freedom leads to fewer calculations per time step, and (2) the increase of the spatial resolution of the system allows larger stable time step in explicit solvers \cite{muller2002coarse}. CG is widely used in the field of atomistic simulations and molecular dynamics \cite{noid2008multiscale-I,noid2008multiscale-II}. CG models can be formulated relatively easily for homogeneous atomistic systems, consisting of a repetitive structure. Furthermore, heterogeneous atomistic models for materials such as protein-based materials, can be coarse grained if they are homogeneous at the meso-scale, which means that the local effects are negligible and they produce nearly homogeneous global behaviors \cite{cranford2010coarse}. There is a  wide literature relevant  to coarse-graining methods to which the reader is referred to for additional information \cite{PhysRevLett.105.237802,praprotnik2005adaptive,rzepiela2011hybrid,brini2013systematic,henderson1974uniqueness,reith2003deriving,PhysRevE.52.3730,ercolessi1994interatomic,izvekov2005multiscale,izvekov2005systematic,mullinax2009generalized}.  


In the study presented in this paper, multiscale homogenization and coarse-graining are combined, and a Homogenization Coarse Graining (HCG) framework is presented. The homogenization algorithm recently developed by Rezakhani and Cusatis \cite{rezakhani2016asymptotic,rezakhani2017adaptive} is employed to obtain the effective response of the FS model with actual particle size as well as the CG model with enlarged particles. These effective responses of the FS and CG RVEs are then used by an automatic parameter identification technique based on the nonlinear least square method to calibrate the CG LDPM parameters. Finally, several numerical examples are performed by the both full FS LDPM and the CG LDPM to verify accuracy and effectiveness of the developed framework.

\section{The Lattice Discrete Particle Model (LDPM)}
\label{ldpm-rev}
\begin{figure}[t!]
	\centering 
	{\includegraphics[width=\textwidth]{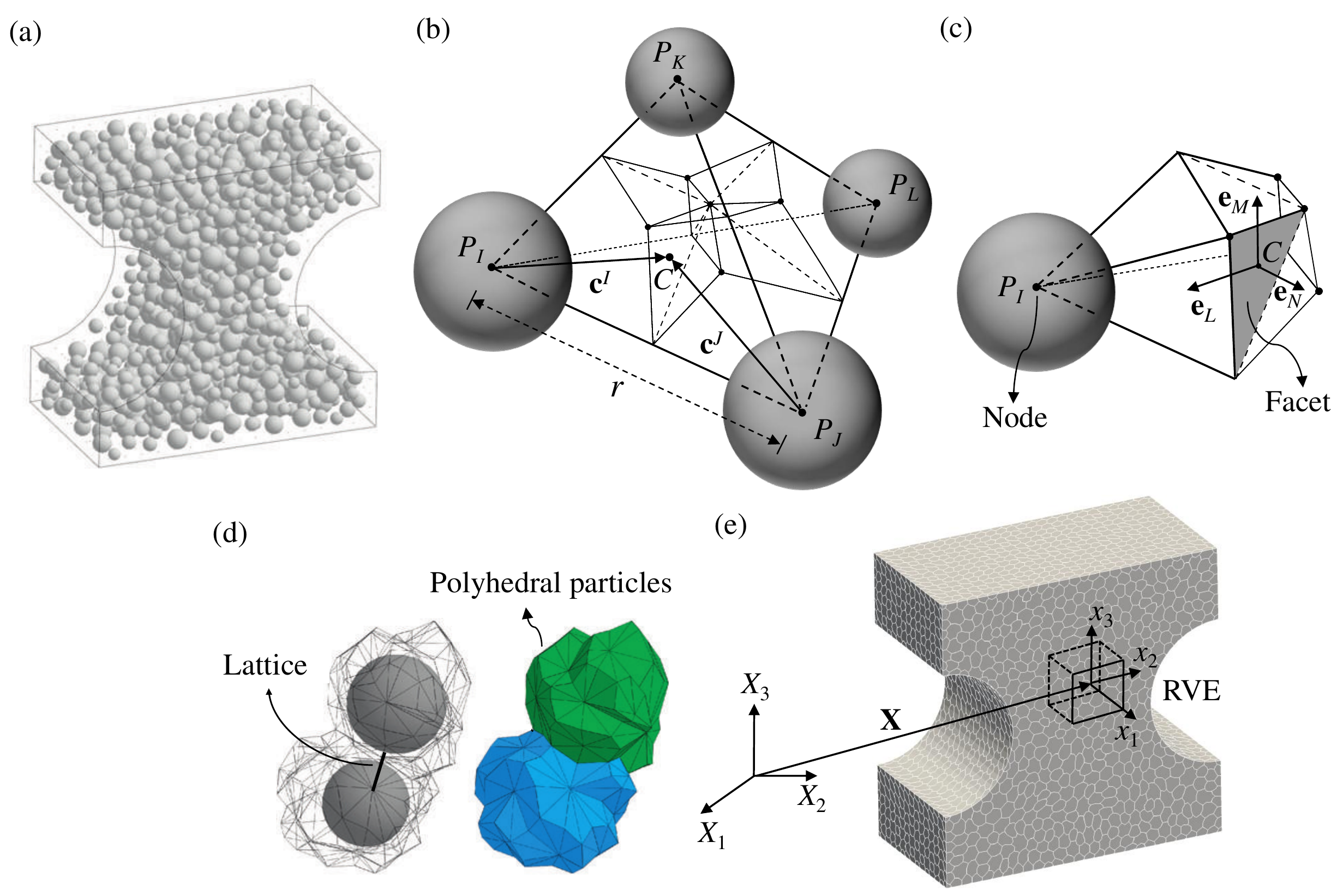}}
	\caption{(a) Spherical aggregate distribution in a dogbone specimen. (b) LDPM tetrahedron connecting four adjacent aggregate particles and its associated tessellation. (c) Tetrahedron portion associated with aggregate $P_I$. (d) Two adjacent LDPM polyhedral cells enclosing the associated aggregate pieces. (e) Polyhedral cell representation of a dogbone specimen.}
	\label{LDPM}
\end{figure}

Starting from the concrete mix design (cement content, $c $; water-to-cement ratio, $w/c$; and aggregate-to-cement ratio, $a/c$. LDPM constructs the geometrical representation of concrete meso-structure through the following steps: (1) The coarse aggregate pieces, whose shapes are assumed to be spherical, are introduced into the concrete volume by a try-and-reject random procedure. Aggregate diameters are determined by sampling an aggregate size distribution function consistent with a Fuller sieve curve: $F(d)=(d/d_a)^n$, where $d=$ particle diameter, $d_a=$ maximum aggregate size. For computational reasons, the Fuller curve is truncated with a certain minimum aggregate size, $d_0$, which defines the resolution of the model. The aggregate distribution in a dogbone specimen obtained with this procedure is depicted in Fig. \ref{LDPM}a. (2) Zero-radius aggregate pieces (nodes) are placed over the external surfaces to facilitate the application of boundary conditions. (3) Delaunay tetrahedralization of the generated aggregate centers and a dual three-dimensional domain tessellation (not identical to the Voronoi tessellation) are carried out to obtain a network of triangular facets as shown in Fig. \ref{LDPM}b. For each particle, combining the relevant tessellation portions from all Delaunay tetrahedral connected to the same node, one obtains a corresponding polyhedral cell which encloses the spherical aggregate. Figure \ref{LDPM}c depicts the portion of the tetrahedral element related to a generic aggregate particle. Two adjacent polyhedral cells interacting through shared triangular facets are depicted in Fig. \ref{LDPM}d. The triangular facets, on which strain and stress quantities are defined in vectorial form, are assumed to be the potential material failure locations. Figure \ref{LDPM}e presents the polyhedral cell representation of the dogbone specimen shown in Fig. \ref{LDPM}a. It is worth pointing out that the spherical aggregate particles are generated to build the discrete system resembling concrete real meso-structure but they are not directly used in the numerical solution procedure. Instead, the centroid of the spherical particles, called ``node'', and the associated polyhedral cells, called simply ``cells'', are the geometrical units that are employed in the numerical analysis. In this paper nodes and cells will be symbolized with $P_I$ and $C_I$, respectively, with $I=1,..., N$, $N=$ total number of aggregate particles.

Reference \cite{cusatis-ldpm-1} provides detailed discussion of the algorithms used for the generation and the placement of the particle system as well as the adopted tessellation.

Three sets of equations are necessary to complete the discrete model framework: definition of strain vector on each facet, constitutive equations which relate facet stress vector to facet strain vector, and cell equilibrium equations.

\subsection{Facet strain definition} 
Rigid body kinematics is employed to describe the deformation of the lattice/cell system, and the displacement jump, $\llbracket \mathbf{u} \rrbracket^{IJ}$, at the centroid of a generic facet shared by two adjacent nodes $P_I$ and $P_J$ (see Fig. \ref{LDPM}b) is used to define measures of strain as 
\begin{equation} 
\label{eps} 
\epsilon_{\beta}=r^{-1} \llbracket \mathbf{u} \rrbracket^{IJ} \cdot \mathbf{e}_{\beta}^{IJ} =r^{-1} \left(\mathbf{U}^J + \mb \Theta^{J} \times \mathbf{c}^{J} - \mathbf{U}^I - \mb{\Theta}^I \times \mathbf{c}^I \right) \cdot \mathbf{e}_{\beta}^{IJ}
\end{equation}
where $\beta=N,M,L$. The strain definition in Eq. \ref{eps} is valid under the assumption of small strains, small displacements, and small rotations. In Eq. \ref{eps} and in the remainder of the paper, the double subscript $IJ$ is included only for quantities that change sign upon change in the $I$ and $J$ order: for examples, $\mathbf{e}_{\beta}^{IJ}=-\mathbf{e}_{\beta}^{JI}$. To avoid shear locking phenomena \cite{cusatis-ldpm-1}, Eq. \ref{eps} is written with reference to the facet projected orthogonally to the line connecting the two relevant aggregate particles as opposed to the original facet. Furthermore, it is worth pointing out that the strain definition in Eq. \ref{eps} is consistent with the projection on the projected facet orientations of the classical strain tensor \cite{cusatis2013high,lale2017isogeometric,cusatis2017discontinuous}. 

In Eq. \ref{eps}, $\epsilon_N$ is the facet normal strain component; $\epsilon_M$ and  $\epsilon_L$ are the facet tangential strain components; $r=|\mathbf{x}^{IJ}|$, $\mathbf{x}^{IJ}=\mathbf{x}^J-\mathbf{x}^I$ is the vector connecting the nodes $P_I$ and $P_J$; $\mathbf{e}_{\beta}^{IJ}$ are unit vectors defining a facet local Cartesian system of reference such that $\mathbf{e}^{IJ}_{N}$ is orthogonal to the projected facet and $r^{-1}\mathbf{e}^{IJ}_{N} \cdot \mathbf{x}^{IJ} =1$, $\mathbf{e}^{IJ}_{M}$ and $\mathbf{e}^{IJ}_{L}$ are tangential to the projected facet; $\mathbf{c}^{I}$ and $\mathbf{c}^{J}$ are vectors connecting the two nodes $I$ and $J$, respectively, to the centroid of the facet, see Fig. \ref{LDPM}b and c for the visualization of these geometrical entities; $\mathbf{U}^I$, $\mathbf{U}^J$ = displacement vectors of node $P_I$ and $P_J$; $\mb{\Theta}^{I}$, $\mb{\Theta}^{J}$ = rotation vectors of node $P_I$ and $P_J$.

\subsection{Facet vectorial constitutive equations} 
Next, a vectorial constitutive law governing the behavior of the material is imposed at the centroid of each facet. In the elastic regime, the normal and shear stresses are proportional to the corresponding strains: $t_{N} = E_N \epsilon_{N}; ~  t_{M} = E_T \epsilon_{M}; ~  t_{L} = E_T \epsilon_{L}$, where $E_N=E_0$; $E_T=\alpha E_0$; $E_0=$ effective normal modulus; $\alpha=$ shear-normal coupling parameter. For stresses and strains beyond the elastic limit, concrete mesoscale nonlinear phenomena are characterized by three mechanisms: (a) mesoscale mixed-mode fracture; (b) pore collapse and material densification; and (c) mesoscale frictional behavior. The corresponding facet level vectorial constitutive equations are briefly described below.

\subsubsection{Mesoscale fracture and cohesion due to tension and tension-shear} 
For tensile loading ($\epsilon_N>0$), the fracturing behavior is formulated \cite{cusatis2003confinement,cusatis2006confinement} through an effective strain, $\epsilon = [\epsilon_N^{2}+\alpha (\epsilon_M^{2} + \epsilon_L^{2})]^{1/2}$, and effective stress, $t = [t _{N}^2+  (t _{M}^2+t _{L}^2) / \alpha]^{1/2}$, which are used to define the facet normal and shear stresses as \mbox{$t _{N}= \epsilon_N(t / \epsilon)$}; \mbox{$t _{M}=\alpha \epsilon_{M}(t / \epsilon)$}; \mbox{$t _{L}=\alpha \epsilon_{L}(t / \epsilon)$}. The effective stress $t$ is incrementally elastic ($\dot{t}=E_0\dot{\epsilon}$) and must satisfy the inequality $0\leq t \leq \sigma _{bt} (\epsilon, \omega) $ where 
\begin{equation}
\sigma_{bt} = \sigma_0(\omega) \exp \left[-H_0(\omega) \frac{\langle \epsilon_{max}-\epsilon_0(\omega) \rangle}{\sigma_0(\omega)}  \right]
\end{equation}    
in which $\langle x \rangle=\max \{x,0\}$; $\epsilon_0(\omega) = \sigma_0(\omega)/E_0$; $\tan(\omega) =\epsilon _N / \sqrt{\alpha} \epsilon_{T}$ = $t_N \sqrt{\alpha} / t_{T}$ in which $\epsilon_T=(\epsilon_M^{2} + \epsilon_L^{2})^{1/2}$ and $t_T=(t_M^{2} + t_L^{2})^{1/2}$. The symbol $\omega$ is the variable that defines the degree of interaction between shear and normal loading; $\epsilon_{max} = (\epsilon_{N,max}^{2}+\alpha \epsilon_{T,max}^{2})^{1/2}$ is a history dependent variable and $\epsilon_{max} = \epsilon$ in the absence of unloading. The post peak softening modulus is defined as $H_{0}(\omega)=H_{t}(2\omega/\pi)^{n_{t}}$, where $n_t=0.2$ is the softening exponent, $H_{t}$ is the softening modulus in pure tension ($\omega=\pi/2$) expressed as $H_{t}=2E_0/\left(\ell_t/r-1\right)$ \cite{bavzant1983crack}; $\ell_t=2E_0G_t/\sigma_t^2$; $\sigma_t$ is the mesoscale tensile strength; and $G_t$ is the mesoscale fracture energy. This formulation provides a smooth transition between pure tension and pure shear ($\omega=0$) with a parabolic variation for strength given by $\sigma_{0}(\omega )=\sigma _{t}r_{st}^2 [-\sin(\omega)+(\sin^2(\omega)+4 \alpha \cos^2(\omega) / r_{st}^2)^{-1/2} ] / [2 \alpha \cos^2(\omega)]$, where $r_{st} = \sigma_s/\sigma_t$ is the mesoscale shear to tensile strength ratio.

\subsubsection{Pore collapse, compaction, and frictional behavior in compression}
For compression ($\epsilon_N<0$) the constitutive equations simulate pore collapse, compaction, and frictional behavior in compression.
Normal stresses for compressive loading are computed through the inequality $-\sigma_{bc}(\epsilon_D, \epsilon_V)\leq t_N \leq 0$, where $\sigma_{bc}$ is a strain-dependent boundary, function of the element volumetric strain, $\epsilon_V$, and the facet deviatoric strain, $\epsilon_D=\epsilon_N-\epsilon_V$. The volumetric strain is computed by the volume variation of the LDPM tetrahedron as $\epsilon_V= \Delta V/ 3V_0$ and is assumed to be constant for all facets belonging to a given tetrahedron. For $-\sigma_{bc}(\epsilon_D, \epsilon_V) < t_N < 0$ the behavior is incrementally elastic: $\dot{t}_N=E_N \dot{\epsilon}_N$.

Beyond the elastic limit, $-\sigma_{bc}$ models pore collapse and compaction/rehardening. $\sigma_{bc}$ is defined as follows:
\begin{equation}
\sigma_{bc} = \left\lbrace
  \begin{matrix}
    \sigma_{c0}  &\textrm{for}& -\epsilon_{V}\leq 0 \\
    \sigma_{c0} + H_c \langle-\epsilon_{V}-\epsilon_{c0}\rangle   &\textrm{for}& 0 \leq -\epsilon_{V}\leq \epsilon_{c0} \\
    \sigma_{c1} \exp \left[H_c( -\epsilon_{V}-\epsilon_{c1} ) /\sigma_{c1} \right] & \textrm{otherwise} &
  \end{matrix}
\right.
\end{equation}
$\sigma_{c0}$ is the mesoscale compressive yielding stress, 
$\epsilon_{c0}=\sigma_{c0}/E_0$ is the compaction strain at the beginning of pore collapse, $\epsilon_{c1}=\kappa_{c0} \epsilon_{c0}$ is the compaction strain when rehardening begins, $\sigma_{c1}=\sigma_{c0}+H_c(\epsilon_{c1}-\epsilon_{c0}) $, $H_c=H_{c1}+\left(H_{c0}-H_{c1}\right)/\left(1+5\langle r_{DV}-1\rangle\right)$, and $r_{DV}=\lvert \epsilon_D\rvert/(\epsilon_{V0}-\epsilon_V)$ for $\epsilon_V\leq0$ and $r_{DV}=\lvert \epsilon_D\rvert/\epsilon_{V0}$ for $\epsilon_V>0$, $\epsilon_{V0}=0.1\epsilon_{c0}$, $H_{c1}=0.1E_0$. $\kappa_{c0}$, and $H_{c0}$ are assumed to be material parameters \cite{ceccato2017simulation}.

The incremental shear stresses are computed by means of a non-associative plastic constitutive equation as  $\dot{t}_M=E_T(\dot{\epsilon}_M-\dot{\epsilon}^{p}_M)$ and \mbox{$\dot{t}_L=E_T(\dot{\epsilon}_L-\dot{\epsilon}^{p}_L)$}; $\dot{\epsilon}_M^{p}=\dot{\epsilon}_L^{p}=0$ in the elastic regime, $\varphi(\sigma_N,\sigma_M,\sigma_L)<0$; \mbox{$\dot{\epsilon}_M^{p}=\dot{\lambda} \partial \psi / \partial t_M$}, \mbox{$\dot{\epsilon}_L^{p}=\dot{\lambda} \partial \psi / \partial t_L$} during plastic flow, $\varphi(t_N,t_M,t_L)=0$. $\varphi(t_N,t_M,t_L)$ is the yielding surface, $\lambda$ is the plastic multiplier, and $\psi(t_M,t_L)=\psi_0(t_M^2+t_L^2)^{1/2}$ is the plastic potential.

The yielding surface is defined as 
\begin{equation}
\varphi=\left(t_M^2+t_L^2 \right)^{1/2} -\sigma_s - \mu_0 \sigma_{N0}\left[1 - \exp \left(\frac{t_N}{\sigma_{N0}}\right)\right]
\end{equation}
where $\sigma_{N0}$ is the transitional normal stress; and $\mu_0$ is the internal friction coefficient. 

For additional details on the constitutive equations the reader may want to consult previous LDPM work \cite{cusatis-ldpm-1,cusatis-ldpm-2,ceccato2017simulation,cusatis-Jovanca}.

\subsection{Cell equilibrium equations} 

Finally, the governing equations of the LDPM framework are completed through the equilibrium equations of each individual cell $C_I$, which read
\begin{equation} \label{motion-1}
\sum_{\mathcal{F}^I} A \mathbf{t}^{IJ} + V^I \mathbf{b}^0 = 0; \hspace{0.5 in} \sum_{\mathcal{F}^I} A \mathbf{c}^I \times \mathbf{t}^{IJ} = 0
\end{equation}
where $\mathcal{F}^I$ is the set of facets that form the cell $C_I$; $A$ is the area of the projected facet; $V^I$ is the volume of the cell ;  $\mathbf{b}^0$ is the body force vector (assumed to be uniform over the volume); $\mathbf{t}^{IJ}= t_{\alpha}\mathbf{e}^{IJ}_{\alpha} = t_N \mathbf{e}^{IJ}_{N} + t_M \mathbf{e}^{IJ}_{M} + t_L \mathbf{e}^{IJ}_{L}$ is the resultant stress vector applied on each triangular facet.

LDPM is implemented in the computational software MARS \cite{mars-1} and has been used successfully to simulate concrete behavior in different types of laboratory experiments \cite{cusatis-ldpm-2}. Furthermore, LDPM has shown superior capabilities in modeling concrete behavior under dynamic loading \cite{cusatis-Jovanca,smith2017numerical}, Alkali Silica Reaction (ASR) deterioration \cite{cusatis-mohammed,alnaggar2017modeling,pathirage2018effect}, failure, and fracture of fiber-reinforced concrete \cite{cusatis-Ed1,cusatis-Ed2,jin2016lattice}. Finally, LDPM was used successfully to simulate ultra-high performance \cite{wan2016analysis,wan2018age}, waterless concrete \cite{wan2016novel}, fiber reinforced polymer (FRP) confined concrete \cite{ceccato2017simulation}, rocks \cite{ashari2017lattice,li2017multiscale}, and to build reduced-order models of failure \cite{ceccato2018proper}. 

\subsection{Identification of model parameters}
\label{sec:identification-ldpm-parameters}
LDPM depends on a number of model parameters that need to be identified by fitting experimental data relevant to the specific concrete to be simulated. The identification can be performed with the following sequence
\begin{enumerate}
\item $E_0$ and $\alpha$ are identified on the basis experimental data relevant to the elastic behavior or with estimates of macroscopic elastic modulus and Poisson ratio. For $\alpha=0.25$ one obtains the usual Poisson's ratio of 0.18. This value will be used in the rest of the paper
\item $\sigma_t$ and $\ell_t$ are identified with experimental data relevant to tensile fracture such as, e.g., three-point bending tests, splitting (Brazilian) tests, and modulus of rupture tests, or with estimates of macroscopic tensile strength and macroscopic fracture energy.
\item $\sigma_{c0}$, $\kappa_{c0}$, and $H_{c0}$ are identified with experimental data relevant to the behavior of concrete under hydrostatic compression. In absence of specific experimental data the following values can be used for normal strength concrete $\sigma_{c0}=100$ MPa, $\kappa_{c0}=4.0$ and $H_{c0}=0.4 E_0$ .
\item $\sigma_s$ (or, equivalently, $r_{st}=\sigma_s/\sigma_t$), $\mu_0$, and $\sigma_{N0}$ are identified with experimental data relevant to triaxial compression tests. If only unconfined compression strength is available, the identification can be restricted to $\sigma_s$ and the values $\mu_0=0.2$, and $\sigma_{N0}=600$ MPa can be used.
\end{enumerate}

\section{Asymptotic Expansion Homogenization}
The homogenization technique formulated by Rezakhani and Cusatis \cite{rezakhani2016asymptotic} is employed in this study to calculate the mechanical response of a Representative Volume Element of LDPM.

Let us consider two macroscopic coordinate systems, $\bf X$ and $\bf x$, that define the position of the centroid of a generic LDPM RVE and the position within the RVE, respectively (Fig. \ref{LDPM}e); in addition let us consider a meso-scale coordinate system $\bf y$. In $\bf X$ and $\bf x$, the material is considered homogeneous and all the material heterogeneity is only visible in $\bf y$. Thanks to the separation of scales hypothesis one can write $\mathbf{x}=\eta \mathbf{y}$ where $\eta$ is a very small positive scalar, $0< \eta <<1$. It is worth noting that, for the separation of scales to hold the RVE size and the LDPM cell size should be much smaller than the size of the sample. For the sake of representation clarity, however, Fig. \ref{LDPM}e violates such requirements.

\subsection{Asymptotic expansion}
For discrete particulate systems such LDPM the asymptotic expansion of the problem variables can be obtained by assuming the existence of two multiscale fields, $\mathbf{u}(\mathbf{x},\mathbf{y})$ and $\boldsymbol{\uptheta}(\mathbf{x},\mathbf{y})$, that coincide with displacements and rotations of a generic node P$_I$ when evaluated for $\mathbf{x}=\mathbf{x}^I$ and $\mathbf{y}=\mathbf{y}^I$:  $\mathbf{U}^I = \mathbf{u}(\mathbf{x}^I, \mathbf{y}^I)$ and ${\boldsymbol \Theta}^I = \boldsymbol{\uptheta}(\mb{x}^I, \mb{y}^I)$. These fields can be approximated with the following asymptotic expansions
\begin{equation}
\mathbf{u}(\mathbf x, \mathbf y) \approx \mathbf u^0(\mathbf x, \mathbf y)+\eta \mathbf u^1(\mathbf x, \mathbf y) 
\label{disp-expansion}
\end{equation}
\begin{equation}
\boldsymbol{\uptheta}  (\mathbf x, \mathbf y)  \approx \eta^{-1} \boldsymbol{\upomega} ^{0}(\mathbf x, \mathbf y) +  \boldsymbol{\upvarphi}^0(\mathbf x, \mathbf y)+  \boldsymbol{\upomega}^{1}(\mathbf x,\mathbf y)+\eta \boldsymbol{\upvarphi}^{1}(\mathbf x, \mathbf y) 
\label{rot-expansion}
\end{equation}
where second order terms and higher are neglected; $\mathbf{u}^0(\mb x, \mb y)$, and $\mathbf{u}^1(\mathbf x, \mathbf y)$ are the coarse- and fine-scale displacement vectors, respectively, which are continuous functions with respect to $\mathbf{x}$ and discrete $\mathbf{y}$. In addition, since the rotation field can be always imagined as the curl of an appropriate displacement vector, it is simple to show \cite{rezakhani2016asymptotic} that $\boldsymbol{\upomega}^{0}$, $\boldsymbol{\upomega}^{1}$ are fine-scale rotations whereas $\boldsymbol{\upvarphi}^{0}$ and $\boldsymbol{\upvarphi}^{1}$ are coarse-scale rotations. 

By introducing the asymptotic expansions of displacements and rotations Eq. \ref{eps} and accounting for the rigid body motion of the RVE \cite{rezakhani2016asymptotic}, one obtains the asymptotic expansion of the facet strains in the form
\begin{equation}
\epsilon_{\alpha}= \epsilon_{\alpha}^0 + \eta \epsilon_{\alpha}^1
\end{equation}
in which, by neglecting the effect of macroscopic curvatures \cite{rezakhani2016asymptotic}, one has
\begin{equation}\label{eps-expansion-zero}
\epsilon_{\alpha}^0 = r^{-1} \llbracket \mathbf{u}_1 \rrbracket^{IJ} \cdot \mathbf{e}_{\alpha}^{IJ} + \mathbf{S}_\alpha  \otimes \boldsymbol{\upvarepsilon} + \mathbf{A}_\alpha  \otimes \boldsymbol{\upxi} = \epsilon_{\alpha}^{f} + \epsilon_{\alpha}^{c}+\xi_{\alpha}^{c}
\end{equation}
The zero-order facet strains, $\epsilon_\alpha^0$, are composed of three terms: the first, $\epsilon_\alpha^f$, are the fine-scale facet strains; the second and third, $\epsilon_{\alpha}^{c}$ and $\xi_{\alpha}^{c}$, are the projections of the coarse-scale symmetric strain tensor, $\boldsymbol{\upvarepsilon}$,  components $\varepsilon_{ij}=(u_{j,i}^0+u_{i,j}^0)/2$, and the coarse-scale anti-symmetric strain tensor, $\boldsymbol{\upxi}$, components $\xi_{ij}=(u_{j,i}^0-u_{i,j}^0)/2-v_{ijk}\omega^0_k$. The tensor $v_{ijk}$ is the Levi-Civita permutation symbol and the projection operators, $\mathbf{S}_\alpha$ and $\mathbf{A}_\alpha$, have components defined as $S^\alpha_{ij} = (e^{IJ}_{Ni} e^{IJ}_{\alpha j}+e^{IJ}_{Nj} e^{IJ}_{\alpha i})/2$ and $A^\alpha_{ij} = (e^{IJ}_{Ni} e^{IJ}_{\alpha j}-e^{IJ}_{Nj} e^{IJ}_{\alpha i})/2$. It is worth pointing out that the fine-scale strain $\epsilon_\alpha^f$ does not coincide with the first order facet strains, $\epsilon_{\alpha}^1$, whose derivation is reported in Ref. \cite{rezakhani2016asymptotic}. 

\subsection{Multiscale equilibrium equations}
By using the asymptotic expansion of the facet strains, one can obtain the asymptotic expansion of the facet stress tractions, which, along with Eq. \ref{motion-1}, allows the derivation of equilibrium equations of order zero (the RVE problem) and of order one (the coarse-scale or macroscopic problem).

The RVE equilibrium equations read 
\begin{equation}\label{RVE-1}
\sum_{\mathcal{F}_I}{{A}\, t^{0}_{\alpha} {\mathbf{e}}_\alpha^{IJ}} = \mathbf{0} \hspace{0.5in} \sum_{\mathcal{F}_I}{A \left(\mathbf{c}^I \times t^{0}_{\alpha} {\mathbf e}_\alpha^{IJ} \right)} = \mathbf{0}  
\end{equation}
where the zero order stress tractions $t^{0}_{\alpha}$ are computed through the constitutive equations reported earlier and with reference to the strains $\epsilon_\alpha^0$. Equation \ref{eps-expansion-zero} can be also rewritten as $\epsilon_{\alpha}^{0}= \epsilon_\alpha^f - (- \epsilon_{\alpha}^{c}-\xi_{\alpha}^{c})$ which interprets the projection of the coarse-scale strains (with negative sign) as facet eigenstrains driving the RVE problem.  Solution of the RVE equilibrium equations under periodic boundary conditions yields the zero order traction, $t_\alpha^0$, which permit the calculation of the coarse-scale, symmetric and anti-symmetric stress tensors:
\begin{equation} \label{macro-stress-formula}
\boldsymbol{\upsigma} = \frac{1}{2V_0} \sum_I \sum_{\mathcal{F}_I}{A} r t^0_\alpha \mathbf{S}_\alpha \hspace{0.25 in} \mathrm{and} \hspace{0.25 in} \boldsymbol{\uptau} = \frac{1}{2V_0} \sum_I \sum_{\mathcal{F}_I}{A} r t^0_\alpha \mathbf{A}_\alpha
\end{equation}
where $V_0$ is the RVE volume. The homogenization procedure provides also the coarse-scale couple tensor, which, however, it was shown to be negligible for the case of LDPM-based constitutive equations \cite{li2017multiscale}.

Finally, averaging the order one equilibrium equations leads to the coarse-scale equilibrium equations \cite{rezakhani2016asymptotic}, which read

\begin{equation} \label{macro-eq-cont}
\nabla^{\mathrm{T}} \cdot \boldsymbol{\upsigma} + \mathbf{b} = \mathbf{0} \hspace{0.25 in} \mathrm{and} \hspace{0.25 in} \boldsymbol{\uptau}=\mathbf{0} 
\end{equation}

Since the antisymmetric stress tensor and the couple tensor are zero, the antisymmetric strain tensor must be zero as well: $\boldsymbol{\upxi}=\mathbf{0}$. This allows performing the numerical implementation of the LDPM homogenization framework without coarse-scale rotational degrees of freedom and with standard displacement-based finite elements. In this study linear tetrahedral elements are employed.

\section{Formulation of the Coarse Graining Scheme} \label{Coarse Graining Scheme}
\begin{figure}[t!]
	\centering 
	{\includegraphics[width=\textwidth]{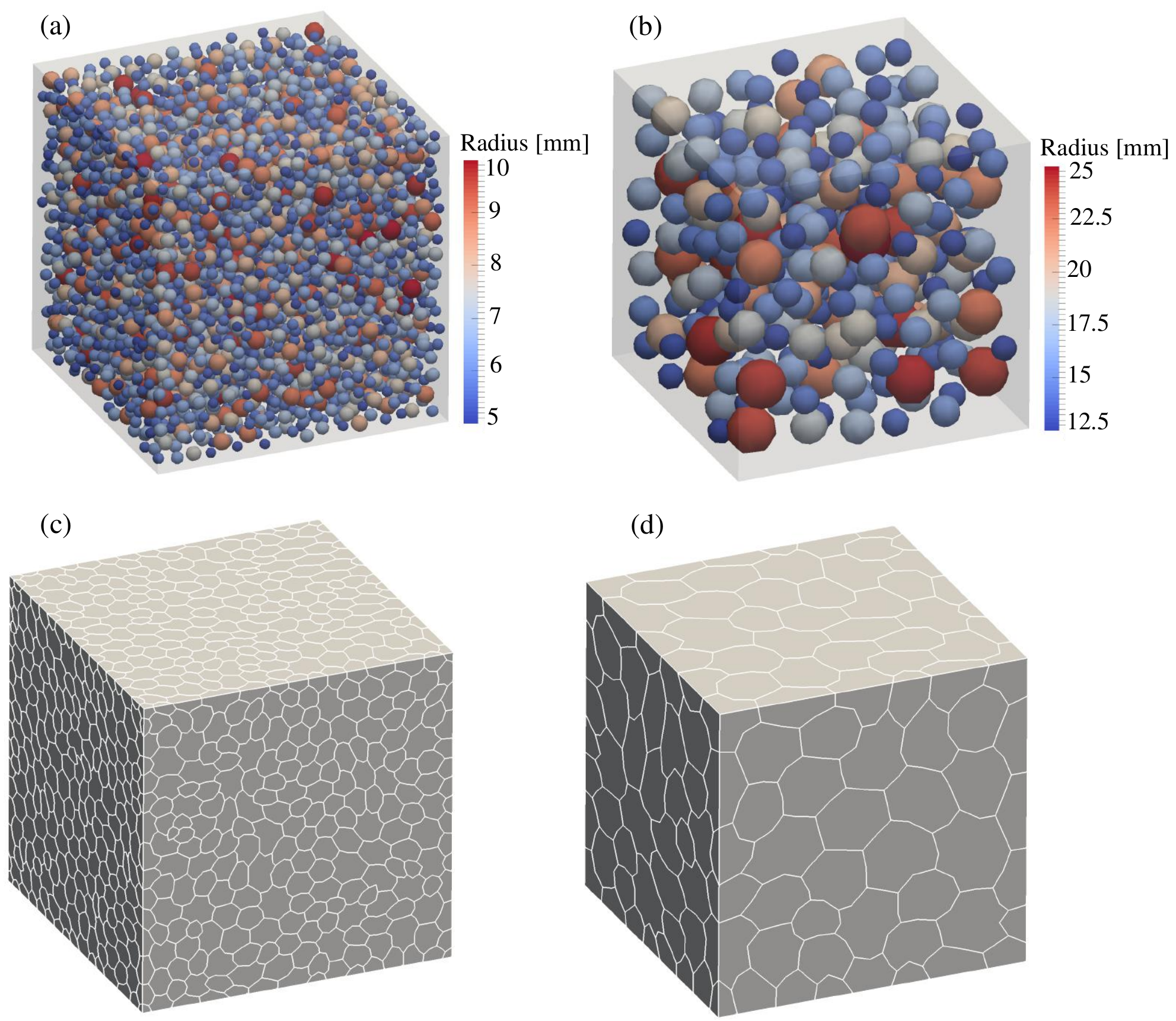}}
	\caption{Particle representation of a generic (a) fine and (b) coarse RVE. Corresponding cell representations of the (c) fine and (d) coarse RVEs.}
	\label{fig:cubemodel}
\end{figure}

The first step of the CG procedure is the selection of a coarsening factor $k_\mathcal{C}$ which provides an appropriate balance between the desired solution accuracy and a reduced computational cost. In the CG system minimum and maximum particle sizes are defined as $d_0^{\mathcal{C}}=k_{\mathcal{C}} d_0$ and $d_a^{\mathcal{C}}=k_{\mathcal{C}} d_a$. Figure \ref{fig:cubemodel} shows the comparison of FS and CG LDPM particle and cell distributions in one RVE of characteristic size $D$ and for $k_{\mathcal{C}}=2.5$. 

If $k_{\mathcal{C}}$ is not excessively large, one can assume that the functional form of the constitutive equations in the coarse system is the same as in the original one. In this case the CG procedure can be reduced to the identification of the parameters of the coarse system. In this study, such identification is carried out with reference to the macroscopic, homogenized response of a certain RVE of material.

For a given set of the LDPM parameters $\mathbf{p}=[E_0~\alpha~\sigma_t~\ell_t~\sigma_{c0}~\kappa_{c0}~H_{c0}~\sigma_s~\mu_0~\sigma_{N0}]^{\mathrm{T}}$; a certain size of the RVE, $D$; and for a generic loading condition the LDPM homogenized response can be computed. One can write
\begin{equation}
\label{eq:homog-response}
\boldsymbol{\upsigma}-\mathbf{f}_D(\boldsymbol{\upvarepsilon},\mathbf{p})=\mathbf{0}
\end{equation}
where $\mathbf{f}_D(\cdot)$ represents the homogenized response for an RVE of size $D$ and it can only be calculated numerically. For a loading history defined in term of strains (deformation control), Eq. \ref{eq:homog-response} can be solved for stresses explicitly. On the contrary, if the loading history is either defined in terms of stresses (load control) or through a combination of strain and stress components (mixed control), the response must be calculated by solving the nonlinear system of equations in Eq. \ref{eq:homog-response}. It is worth pointing out that, in the nonlinear regime  the solution of such system might not exist for particular stress histories due to the strain softening behavior.

Similarly, one has
\begin{equation}
\label{eq:homog-response-coarse}
\boldsymbol{\upsigma}-\mathbf{f}^\mathcal{C}_\mathcal{D}(\boldsymbol{\upvarepsilon},\mathbf{p}^\mathcal{C})=\mathbf{0}
\end{equation}
where $\mathbf{f}^\mathcal{C}_\mathcal{D}(\cdot)$ represents the homogenized response for a coarse RVE of size $\mathcal{D}$ and $\mathbf{p}^\mathcal{C}=[E_0^\mathcal{C}~\alpha^\mathcal{C}~\sigma_t^\mathcal{C}~\ell_t^\mathcal{C}~\sigma_{c0}^\mathcal{C}~\kappa_{c0}^\mathcal{C}~H_{c0}^\mathcal{C}~\sigma_s^\mathcal{C}~\mu_0^\mathcal{C}~\sigma_{N0}^\mathcal{C}]^{\mathrm{T}}$ is the LDPM parameter set for the coarse system.

The imposed stress and strain components (input) for a certain loading condition can be collected in two vectors $\boldsymbol{\upsigma}_{in}$ and  $\boldsymbol{\upvarepsilon}_{in}$, respectively. Similarly, the corresponding  stress and strain components calculated through the solution of the RVE problem can be collected in two other vectors $\boldsymbol{\upsigma}_{out}$ and  $\boldsymbol{\upvarepsilon}_{out}$. In this case, Eq. \ref{eq:homog-response} can be rewritten as $\boldsymbol{\upsigma}_{out}=\mathbf{g}_D(\boldsymbol{\upvarepsilon}_{in}, \boldsymbol{\upsigma}_{in},\mathbf{p})$ and $\boldsymbol{\upvarepsilon}_{out}=\mathbf{h}_D(\boldsymbol{\upvarepsilon}_{in}, \boldsymbol{\upsigma}_{in},\mathbf{p})$.
Similarly, for the CG system one has $\boldsymbol{\upsigma}_{out}^\mathcal{C}=\mathbf{g}^\mathcal{C}_\mathcal{D}(\boldsymbol{\upvarepsilon}_{in}, \boldsymbol{\upsigma}_{in},\mathbf{p}^\mathcal{C})$ and $\boldsymbol{\upvarepsilon}^\mathcal{C}_{out}=\mathbf{h}^\mathcal{C}_\mathcal{D}(\boldsymbol{\upvarepsilon}_{in}, \boldsymbol{\upsigma}_{in},\mathbf{p}^\mathcal{C})$
%

The CG parameters $\mathbf{p}^\mathcal{C}$ can be identified by minimizing the difference between the macroscopic behavior of the CG and FS responses. This can be done efficiently as follows. First the FS solution is calculated for $n=1,...,N$ loading histories characterized by stresses $\boldsymbol{\upsigma}^{nm}_{in}$ and strains $\boldsymbol{\upvarepsilon}^{nm}_{in}$. Each loading history is discretized in $m=1,..., M_n$ increments and the fine scale solution is calculated for $l=1,..., L_n$ RVEs with different mesostructure (different particle configurations). The average FS response can be computed as
\begin{equation}
\label{eq:rves-fine-1}
\boldsymbol{\upSigma}_D^{nm}=\frac{1}{L_n}\sum_{l=1}^{L_n} \mathbf{g}_D(\boldsymbol{\upsigma}_{in}^{nm},\boldsymbol{\upvarepsilon}_{in}^{nm},\mathbf{p})^l
\end{equation}
and
\begin{equation}
\label{eq:rves-fine-2}
\boldsymbol{\upPsi}_D^{nm}=\frac{1}{L_n}\sum_{l=1}^{L_n} \mathbf{h}_D(\boldsymbol{\upsigma}_{in}^{nm},\boldsymbol{\upvarepsilon}_{in}^{nm},\mathbf{p})^l
\end{equation}

This set of FS solutions is the target for the minimization problem. Similarly, a CG approximation of the FS solution can be calculated for a certain set of CG parameters as
\begin{equation}
\label{eq:rves-coarse-1}
\mathbf{G}_{\mathcal{D}}^{nm}(\mathbf{p}^\mathcal{C})=\frac{1}{K_n}\sum_{k=1}^{K_n} \mathbf{g}_{\mathcal{D}}^{\mathcal{C}}(\boldsymbol{\upsigma}_{in}^{nm},\boldsymbol{\upvarepsilon}_{in}^{nm},\mathbf{p}^\mathcal{C})^k
\end{equation}
and
\begin{equation}
\label{eq:rves-coarse-2}
\mathbf{H}_{\mathcal{D}}^{nm}(\mathbf{p}^\mathcal{C})=\frac{1}{K_n}\sum_{k=1}^{K_n} \mathbf{h}_{\mathcal{D}}^{\mathcal{C}}(\boldsymbol{\upsigma}_{in}^{nm},\boldsymbol{\upvarepsilon}_{in}^{nm},\mathbf{p}^\mathcal{C})^k
\end{equation}
for $K_n$ CG RVEs with different particle configurations.

The minimization problem can be then formulated with an objective function obtained by computing the difference between the CG and the FS (target) response and by concatenating the various loading histories. One can write


\begin{equation}
\label{eq:minimization}
\min_{\mathbf{p}^\mathcal{C}} \left[ \sum_{n=1}^N \gamma^n  \sum_{m=1}^{M_n}  \kappa_\Sigma \left( \frac{ \big|\big|\mathbf{G}_{\mathcal{D}}^{nm}(\mathbf{p}^\mathcal{C})-\boldsymbol{\upSigma}_D^{nm}\big|\big| }{\Sigma_n}\right)^2 + \kappa_\Psi \left(\frac{ \big|\big|\mathbf{H}_{\mathcal{D}}^{nm}(\mathbf{p}^\mathcal{C})-\boldsymbol{\upPsi}_D^{nm}\big|\big|}{\Psi_n} \right)^2 \right]^{1/2}
\end{equation}
where $||\mathbf{a}||=(\sum_{kl}a_{kl}^2)^{1/2}$ is the $L^2-$norm of the tensor $\mathbf{a}$; $\Sigma_n=\max_{m}||\boldsymbol{\upSigma}_D^{nm}||$ and  $\Psi_n=\max_{m}||\boldsymbol{\upPsi}_D^{nm}||$ are normalization factors defined for each loading history for stresses and strains, respectively; $\gamma^n$ is a weighting factor for each loading condition defined such that $\sum_n\gamma^n=1$ and $\kappa_\Sigma$, $\kappa_E$ are weighting factors for stresses and strains, respectively ($\kappa_\Sigma+\kappa_E=1$). 
Since the minimization problem in Eq. \ref{eq:minimization} is highly nonlinear, an appropriate nonlinear optimization tool needs to be employed for its solution. Several nonlinear optimization methods are available in literature such as the nonlinear least square method \cite{Dennis}, Kalman filters \cite{Bolzon}, and the ones using artificial intelligence techniques such as Neural networks and Genetic algorithms \cite{Aguir}. The nonlinear least square method, which is simple yet effective and widely used in a broad range of optimization applications, is used in the current study.

The minimization problem in Eq. \ref{eq:minimization} requires the definition of the fine and coarse RVE size and appropriate loading histories. 

In previous work on homogenization of LDPM, \cite{rezakhani2016asymptotic} it was demonstrated that for an RVE size greater than five times the maximum particle size, the homogenized response converges quickly towards an asymptotic response and is insensitive of the particle positions within the RVE only if the macroscopic response is elastic or strain hardening. On the contrary,  in the case of strain softening, the response is always RVE size dependent and it is more brittle as the RVE size increases. This is, of course, not surprising due to damage localization occurring during softening \cite{bazant1997fracture}. 
Also, a certain critical RVE size, $D_{cr}$, exists and it corresponds to a stress strain curve with vertical drop. $D_{cr}$ is directly related to the so-called macroscopic characteristic length \cite{bazant1997fracture}
, $\ell_{ch}$, and for $D>D_{cr}$ the stress strain curve features snap back behavior.
Because of the size dependence of the softening response, the RVE size of the CG system must be assumed equal to the RVE size used to compute the FS (target) response. In this study, the RVE size is taken as $D=\mathcal{D}=6d_a^{\mathcal{C}}$. 

Since the snap back behavior is unstable under both deformation and load control the following condition $D<D_{cr}$ must also be enforced. In turn, this limits the value of the coarse graining factor to the following value $k_\mathcal{C}<0.2D_{cr}/d_a$. 

The loading conditions that need to be selected for the identification as well as the weighting factors for each condition depend on the macroscopic application to be solved and the relevance or not of certain LDPM parameters to that specific application. In the most general case in which all the LDPM parameters are equally important, then the loading conditions can be selected on the basis of typical experimental data sets used for LDPM calibration (see Sec. \ref{sec:identification-ldpm-parameters}).

For the identification of $E_0^\mathcal{C}$ and $\alpha^\mathcal{C}$ the elastic portion of any loading condition can be used. However, since the LDPM macroscopic elastic response is basically independent on the particle size, it is also accurate to assume $E_0^\mathcal{C} = E_0$ and $\alpha^\mathcal{C} = \alpha$.

For $\sigma_t^\mathcal{C}$ and $\ell_t^\mathcal{C}$ a mixed control, uniaxial stress loading path in tension can be used, in this case ($n=1$) one has: e.g., $\varepsilon_{33}^{1m}=\varepsilon_0 \lambda_m$ where $\lambda_m$ is a discretization of the interval 0 to 1; $\sigma_{11}^{1m}=\sigma_{22}^{1m}=0$; $\sigma_{ij}^{1m}=0$ for $i\neq j$;
\begin{equation}
\boldsymbol{\upsigma}_{in}^{1m}=[\sigma_{11}^{1m}~\sigma_{22}^{1m}~\sigma_{12}^{1m}~\sigma_{13}^{1m}~\sigma_{23}^{1m}]^{\mathrm{T}}~~~\boldsymbol{\upsigma}_{out}^{1m}=[\sigma_{33}]
\end{equation}
and
\begin{equation}
\boldsymbol{\upvarepsilon}_{in}^{1m}=[\varepsilon_{33}]~~~ \boldsymbol{\upvarepsilon}_{out}^{1m}=[\varepsilon_{11}^{1m}~\varepsilon_{22}^{1m}~\varepsilon_{12}^{1m}~\varepsilon_{13}^{1m}~\varepsilon_{23}^{1m}]^{\mathrm{T}}
\end{equation}
%
For $\sigma_{c0}^\mathcal{C}$, $\kappa_{c0}^\mathcal{C}$, and $H_{c0}^\mathcal{C}$ a hydrostatic loading path (either deformation or load controlled) can be used. For a hydrostatic load controlled loading path ($n=2$) one has: $\sigma_{11}^{2m}=\sigma_{22}^{2m}=\sigma_{33}^{2m}=p_0\lambda_m$; $\sigma_{ij}^{2m}=0$ for $i\neq j$; $\boldsymbol{\upsigma}_{in}^{2m}=\boldsymbol{\upsigma}^{2m}$; and 
$\boldsymbol{\upvarepsilon}_{out}^{2m}=\boldsymbol{\upvarepsilon}^{2m}$. Finally, for $\sigma_s^\mathcal{C}$, $\mu_0^\mathcal{C}$, and $\sigma_{N0}^\mathcal{C}$, a set of triaxial loading paths in compression can be used. Typically three loading paths are sufficient, namely unconfined compression, confined compression at low confinement and confined compression at high confinement. Since for low confinement the response features strain softening, a mixed control is required. In this case $n$=3,4,5, one can write $\sigma_{11}^{nm}=\sigma_{22}^{nm}=p_n\lambda_m$ for $\lambda_m \leq\lambda_0$ and $\sigma_{11}^{nm}=\sigma_{22}^{nm}=p_n$ for $\lambda_m>\lambda_0$, $p_3$=0 (unconfined compression), $p_4=$ low confining pressure, $p_5=$ high confining pressure;  $\sigma_{kl}^{nm}=0$ for $k \neq l$. In addition,  $\varepsilon_{33}^{nm}=\varepsilon_0 \psi(\lambda)$ for $\lambda_m \leq\lambda_0$ and $\varepsilon_{33}^{nm}=\varepsilon_0 (\lambda_m-\lambda_0)$ for $\lambda_m>\lambda_0$. The function $\psi(\lambda)$ -- characterized by the limits $\psi(0)=0$, $\psi(\lambda_0)=1$-- and the strain $\varepsilon_0$ are defined such that for $\lambda_n \leq \lambda_0$ the strain history for $\varepsilon_{33}^{nm}$ coincides with the one in a hydrostatic test. One obtains
\begin{equation}
\boldsymbol{\upsigma}_{in}^{nm}=[\sigma_{11}^{nm}~\sigma_{22}^{nm}~\sigma_{12}^{nm}~\sigma_{13}^{nm}~\sigma_{23}^{nm}]^{\mathrm{T}}~~~\boldsymbol{\upsigma}_{out}^{nm}=[\sigma_{33}]~~~(n=3,4,5)
\end{equation}
and
\begin{equation}
\boldsymbol{\upvarepsilon}_{in}^{nm}=[\varepsilon_{33}]~~~ \boldsymbol{\upvarepsilon}_{out}^{nm}=[\varepsilon_{11}^{nm}~\varepsilon_{22}^{nm}~\varepsilon_{12}^{nm}~\varepsilon_{13}^{nm}~\varepsilon_{23}^{nm}]^{\mathrm{T}}~~~(n=3,4,5)
\end{equation}
{\emph{Remark}}. The CG procedure introduced above requires knowledge of the fine scale LDPM parameters, which, in turn, must be identified by fitting relevant experimental data. Hence, one might wonder whether it is possible to identify directly the CG LDPM parameters from the experimental data. This strategy avoids the identification of the FS parameters but provides accurate results only for cases in which the CG maximum aggregate size is small enough to simulate with sufficient resolution the geometry of the samples relevant to the experimental data \citep{Smith201413}. Most practical situations do not satisfy this condition. As an example, let's consider a standard concrete mix with a maximum aggregate size of 20 mm for which the compressive strength is measured by testing typical cylindrical samples with diameter equal to 150 mm and length equal to 300 mm. CG simulations with satisfactory resolution of such samples require the CG maximum aggregate size to be smaller than 1/5 of the sample diameter. This clearly limits the coarsening factor to 1.5. For larger coarsening factors the CG system is clearly too coarse for the CG parameters to be identified directly from the experimental data.

\section{Numerical Examples}
This section presents the application of the coarse graining technique discussed above for the simulations of reinforced concrete structures. 

\subsection{Fine scale and coarse scale parameters}
In this type of applications significant multiaxial confinement is not expected and, for this reason, the LDPM parameters governing confined compression have a negligible effect on the structural response. Hence, the CG system may be formulated assuming: $\sigma_{c0}^\mathcal{C}=\sigma_{c0}$, $H_{c0}^\mathcal{C}=H_{c0}$, $\kappa_{c0}^\mathcal{C}=\kappa_{c0}$, and $\mu_0^\mathcal{C}=\mu_0$, and $\sigma_{N0}^\mathcal{C}=\sigma_{N0}$. In this case the CG parameters that need to be identified are $\sigma_t^\mathcal{C}$, $\ell_t^\mathcal{C}$ and $\sigma_s^\mathcal{C}$. 

\begin {table}[t]
\small
\begin{center}
\caption {Fine scale parameters for three different concrete mixes.} \label{tab:fine-scale-parameters}
\begin{tabular} {|c|cccccc|c|ccccc|}
\hline
 & $c$ & $w/c$& $a/c$ & $d_0$ & $d_a$ & $n_F$& $f_c^\prime$ & $E_0$ & $\sigma_t$  & $\ell_t$ & $r_{st}$ & $\sigma_{c0}$ 	\\
  & [kg/m$^3$] & [-] & [-] & [mm] & [mm] & [-] & [MPa] & [MPa] & [MPa] & [mm] & [-] & [MPa] 	\\
\hline 
C30 & 311 & 0.51 & 6.4 & 10 & 20 & 0.58 & 30 & 38,580  &	2.5 & 300 & 4.1 & 100 \\
C35& 311 & 0.51 & 6.4& 10 & 20 & 0.58 & 35& 38,580  &	2.9 & 300 &	4.1 & 100 \\	
C40& 264 & 0.55 & 7.1 &5 & 10 & 0.50 & 43 & 49,958 & 	2.5 & 175 & 5.4 & 150 \\	
\hline
\end{tabular}
\end{center}
\end{table}

The structural applications presented in the following sections are relevant to three concrete mix designs, labeled as C30, C35 and C43, and reported in Table \ref{tab:fine-scale-parameters}. C30 and C35 correspond to the concrete mixes reported in Ref. \cite{kosasize} and whose experiments are simulated in Sec. \ref{sec:deep-beams}. C40 coincides with the concrete mix used in Ref. \cite{duong2006seismic} and whose experiments are analyzed in Sec. \ref{sec:frame}.

Table \ref{tab:fine-scale-parameters} reports also the associated macroscopic compressive strengths and LDPM parameters. For the LDPM parameters not included in Table \ref{tab:fine-scale-parameters}, the default values previously introduced in Sec. \ref{sec:identification-ldpm-parameters} were used in the calculations. It is worth noting that the only available mechanical property for these mix designs was the compressive strength. For this reason the identification of the LDPM parameters was performed as follows: (a) an estimate for the normal modulus was obtained through the formula $E_0=E(4+\alpha)/(2+3\alpha)$ where the macroscopic Young modulus, $E$, was estimated by means of the compressive strength $E=4700 (f^\prime_c)^{1/2}$ with $f^\prime_c$ in MPa \cite{aci2014building}. For the C40 this value was fine tuned to match a peak strain of 2.31$\times 10^{-3}$ which was provided in the experiments; (b) $\ell_t$, $r_{st}$, and $\sigma_{c0}$ were estimated on the basis of previous LDPM work \citep{cusatis-ldpm-2}; and (c) $\sigma_t$ was calibrated to match the macroscopic compressive strength $f_c^{\prime}$ by simulating the unconfined compressive response of cylinders with 150 mm diameter and 300 mm length. The numerically calculated compressive strength was obtained through the average response of three samples with different meso-structural configurations. 

The loading conditions required to identify the coarse scale parameters are simply direct tension ($n=1$ in Sec. \ref{Coarse Graining Scheme}) and unconfined compression ($n=3$ in Sec. \ref{Coarse Graining Scheme}). The objective function in Eq. \ref{eq:minimization} can be then calculated with $\gamma^1=\gamma^3=0.5$ and $\gamma^2=\gamma^4=\gamma^5=0$. In addition, for the examples discussed in this section (a) only the computed stresses were used for the definition of the objective function: $\kappa_\Sigma=1$ and $\kappa_\Psi=0$; (b) each loading history was discretized with an equal number of intervals: $M_1=M_3=1,000$; (c) the FS (target) response was computed with 2 RVEs ($L_1=L_3=2$ in Eqs. \ref{eq:rves-fine-1} and \ref{eq:rves-fine-2}); and (d) the coarse scale solution was computed with the average of 3 RVEs ($K_1=K_3=3$ in Eqs. \ref{eq:rves-coarse-1} and  \ref{eq:rves-coarse-2}). The RVE solutions of the FS response and the approximated CG response were computed by implementing the homogenization procedure as the constitutive equation of a constant strain tetrahedron and by subjecting the tetrahedron to unconfined, uniaxial loading conditions in tension and compression.

\begin{figure}[h!]
\centering
\includegraphics[width=\textwidth]{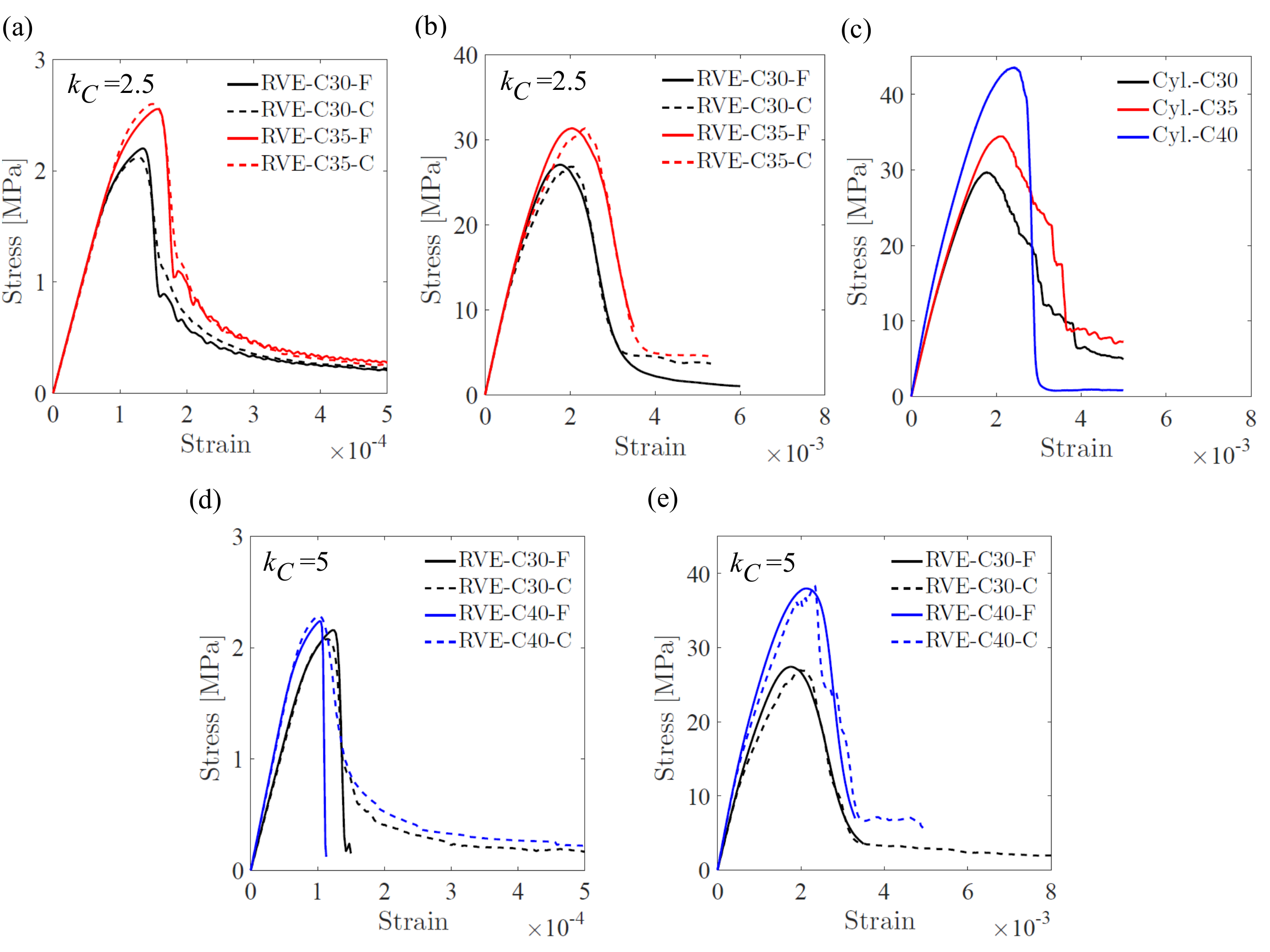}
\caption{Stress vs. strain curves for a) tension with $k_\mathcal{C}=2.5$, b) compression with $k_\mathcal{C}=2.5$, c) Unaxial compression simulation on cylindrical specimen d) tension with $k_\mathcal{C}=5$, e) compression with $k_\mathcal{C}=5$. }
\label{fig:CG-stress-strain}
\end{figure}

With the proposed procedure, the identification of the CG parameters are computed with a reasonable accuracy after 6-7 iterations of minimization scheme. The obtained values for $k_\mathcal{C} =$ 2.5 (for C30 and C35) and 5 (for C30 and C40) are reported in Table \ref{tab:static_parameters_CG}. The corresponding homogenized stress-strain curves obtained from the solution of the FS RVEs and the CG ones with the optimized parameters are plotted in Fig. \ref{fig:CG-stress-strain} for both tension and compression loading conditions. One can see that the results obtained from the CG models are in good agreement with the solution of the FS ones.
\begin {table}[t]
\small
\begin{center}
\caption {Optimized LDPM parameters for coarse grained models.} \label{tab:static_parameters_CG}
\begin{tabular} {|l|cccc|}
\hline
 & C30  & C35 & C30  & C40	\\
\hline 
$k_\mathcal{C}$ [-] & 2.5 & 2.5 & 5 & 5	\\            
$\sigma_t^\mathcal{C}$ [MPa]			& 2.71  	& 3.33		& 2.96	 		&3.05		\\		
$r_{st}^\mathcal{C}$ [-]			& 4.96 	& 4.61 	& 5.10	   		&6.14		\\
$\ell_t^\mathcal{C}$ [mm]			& 302.05  	& 228.71  	& 373.07 		&246.43	\\
\hline
\end{tabular}
\end{center} 
\end{table}

In addition, Fig. \ref{fig:CG-stress-strain} reports, for comparison, the stress vs. strain curves in compression obtained in the simulation of the actual experimental cylinders.

\subsection{Reinforced concrete deep beams}
\label{sec:deep-beams}
In this section the CG procedure formulated above is applied to the simulation of reinforced concrete deep beams that are characterized by small span-to-depth ratio (usually 2 to 4).
%
Many civil engineering applications (transfer beams, foundation pile caps, and coupling beams) feature these structural elements whose failure behavior is rather complex and it involves a variety of failure modes, including diagonal tensile cracking (diagonal splitting), shear compression failure, diagonal compressive failure, and flexural compression failure \cite{ismail2016shear}. Because of such complexity standard analysis techniques fail in most situations to provide satisfactory results especially with reference to size-effect, that is the dependence of structural strength upon size \citep{bazant1984size,bazant1987size,bazant1991size,bazant1997fracture,frosch2017unified,yu2016comparison}.

As far as design is concerned, the Strut and Tie Model (STM) is widely used for deep beams. However, STM has several limitations because (1) it is unable to predict the failure modes which, instead, need to be assumed \cite{tan1999shear}, (2) it requires the definition of an idealized truss configuration which is not obvious in many cases, and (3) does not include size-effect.

The simulation results presented hereinafter are relevant to a comprehensive experimental study carried out by Kosa et al. \cite{kosasize}. They tested deep beams, labeled as B2, B8, B10, B11, B12, and B13, under four-point bending loading condition. Table \ref{tab:beam_dimension} reports the geometrical properties of these beams and it includes length, $L$; depth, $h$; width, $b$; effective depth, $d$; shear span to depth ratio, $a/d$, where the shear span, $a$, is the distance between the center of each support to the center of the first loading point; and the distance between the two loading points, $b_s$. The same table also reports concrete strength $f_c^\prime$ measured for the specific batches of each beam, the reference concrete (C30 or C35) used in the simulations and the adopted coarsening factor $k_\mathcal{C}$. 

In addition, one can find in Table \ref{tab:beam_reinforcement} number of bars, bar type, bar nominal area and yielding stress for the tensile and compressive reinforcements as well as stirrups. For all steel reinforcement the modulus of elasticity was assumed to be 200 GPa and the  hardening modulus 1,500 MPa. All beams had shear reinforcement ($\phi 10_J$ with 75mm spacing) in the central part and at the supports but only beams B8, B11, and B12 had stirrups in the shear span.
\begin{table}[h]
\small
	\centering
	\caption {Deep beams characteristics.} \label{tab:beam_dimension}	
	\begin{tabular}{|l|ccccccccc|}
		\hline
Beam ID        &$L$ [mm]& $h$ [mm]& $b$ [mm]& $d$ [mm]&$a/d$ & $b_s$ [mm] & $f_c'$ [MPa]  & Concrete & $k_\mathcal{C}$ \\        \hline
		B2         & 1,100 & 475   & 240   & 400   &0.5  & 100 & 36.2  & C35 & 2.5\\
		B8         & 1,500 & 475   & 240   & 400   &1.0  & 100 & 37.8  & C35 & 2.5\\
		B10, B11   & 1,900 & 475   & 240   & 400   &1.5  & 100 & 29.2  & C30 & 2.5\\ 
		B12        & 1,900 & 475   & 240   & 400   &1.5  & 100 & 31.3  & C30 & 2.5\\
		B13        & 3,800 & 905   & 480   & 800   &1.5  & 200 & 31.6  & C30 & 5\\
		\hline    
	\end{tabular}
\end{table}
\begin{table}[h!]
\small
\begin{center}
\caption {Reinforcement details of the deep beams. The label ``$\phi \#_{J}$'' denotes the rebar type according to the Japanese standard. The number after the symbol ``@'' denotes the stirrup spacing in mm. The numbers in parenthesis are the nominal cross sectional area for one rebar in mm$^2$ and the yielding stress in MPa.} \label{tab:beam_reinforcement}
	\begin{tabular}{|l|c|c|c|c|}
	\hline
	Beam ID & Tension bars & Compression bars & Shear span stirrups  \\
	\hline
	B2, B10 & 5 $\phi 22_{J}$ (380, 376) & 2 $\phi 10_{J}$ (79, 376) & -  \\
	B11 & 5 $\phi 22_{J}$ (380, 376) & 2 $\phi 10_{J}$ (79, 376) & $\phi 6_{J}$ @65 (29, 376) \\
	B8, B12 & 5 $\phi 22_{J}$ (380, 376) & 2 $\phi 10_{J}$ (79, 376) & $\phi 10_{J}$ @75 (79, 376) \\
	B13 & 10 $\phi 32_{J}$ (804, 398) & 2 $\phi 13_{J}$ (133, 398) & -  \\
	\hline    
	\end{tabular}
\end{center}
\end{table}

The numerical simulations were performed with both the fine and coarse grained LDPM. The steel reinforcement was modeled by means of beam elements governed by classical J2-plasticity with isotropic hardening \citep{jirasek2002inelastic}. Penalty constraint between the LDPM tetrahedra the reinforcement beam elements simulated perfect bond between concrete and steel \citep{Alnaggar_PHD}.
It is worth observing that the assumption of perfect bond is expected to be accurate only up to failure but not in the post-failure regime. Elastic hexahedral solid elements with elastic modulus equal to 200 GPa and Poisson's ratio of 0.3 were used to simulate loading and support plates. Finally, a frictional penalty contact algorithm with friction coefficient equal to 0.5 was used to simulate the interaction between LDPM and the loading plates.  

\begin{figure} [h!]
\centering
\includegraphics[width=\textwidth]{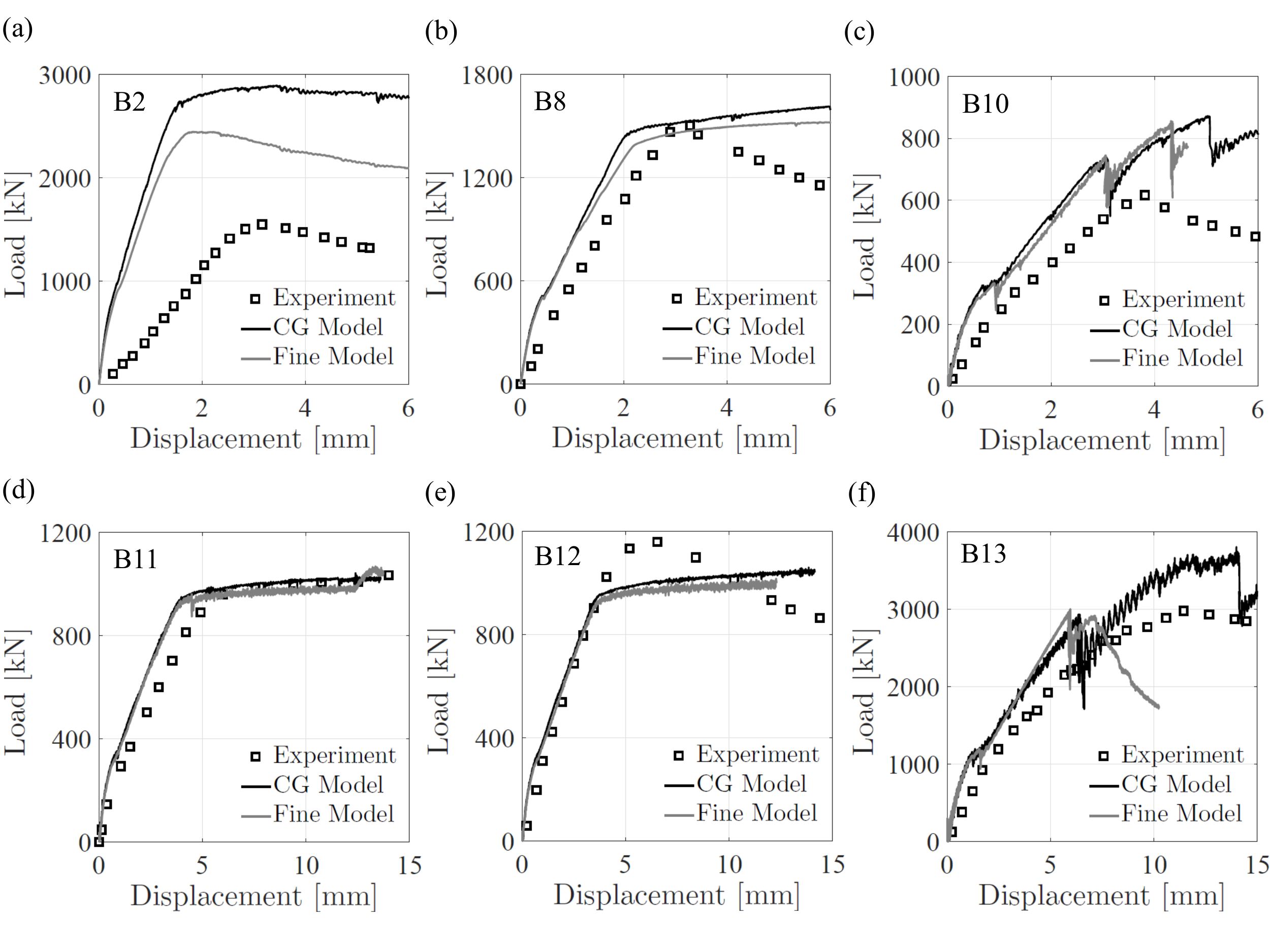}
\caption{Load displacement curves for deep beams.}
\label{fig:LD_DBEAM}
\end{figure}

\begin{figure}[h!]
\centering 
{\includegraphics[width=0.9\textwidth]{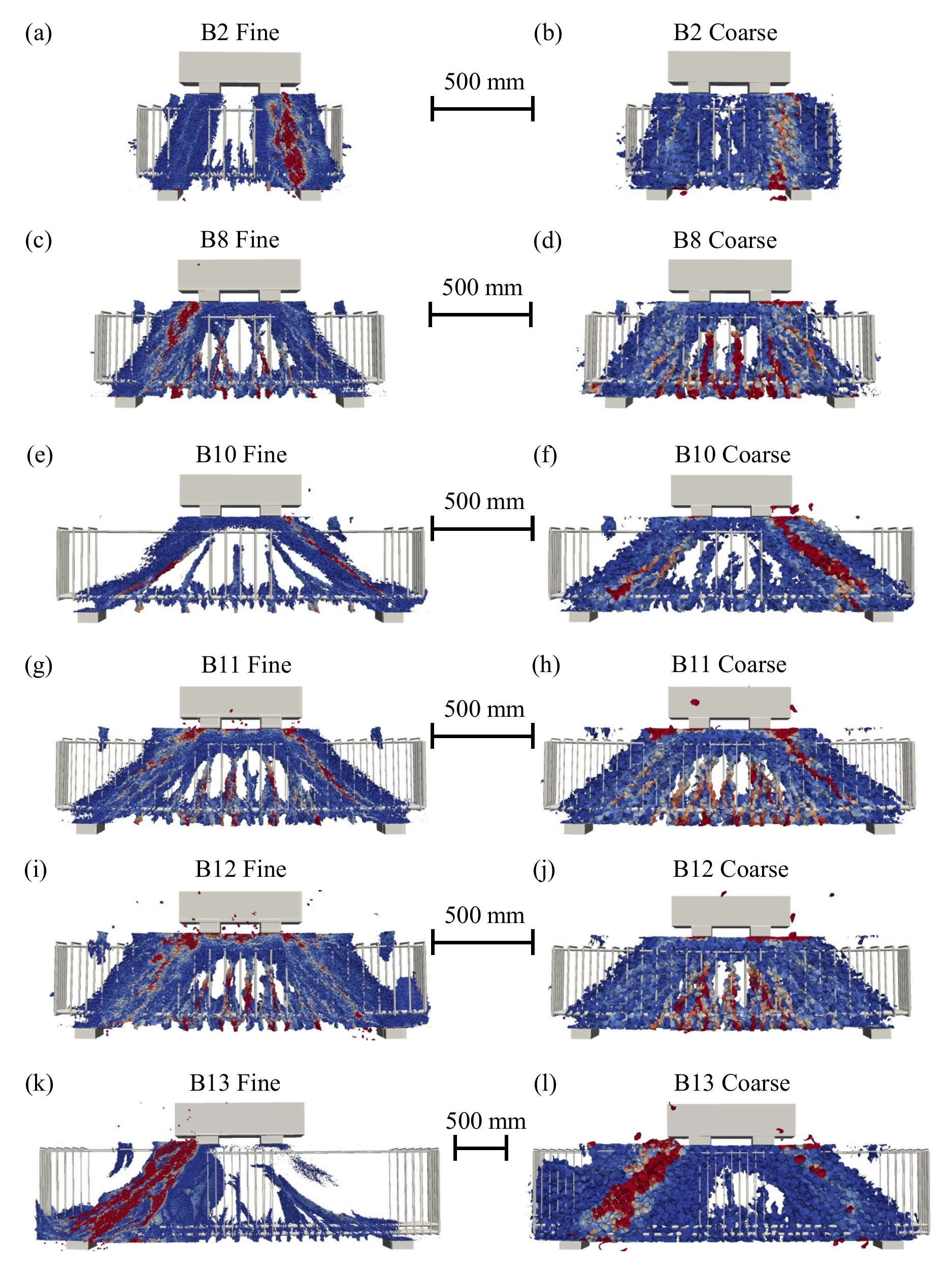}}
\caption{Crack patterns of deep beams.}
\label{fig:CO_DBEAM}
\end{figure}

Figures \ref{fig:LD_DBEAM} and \ref{fig:CO_DBEAM} present the numerical results in terms of load versus midspan deflection curves and crack patterns, respectively. In addition, Table \ref{tab:LD_comparison} reports the peak load for the experiments ($P_0^{ex}$), the FS model ($P^{exp}_0$), and the CG model ($P_0^\mathcal{C}$); as well as the CG error ($err^\mathcal{C}$). 

The results demonstrate that, with exception of beam B2, the predictions of the CG model are in good agreement with the predictions of the FS model in terms of elastic behavior, nonlinear behavior before failure and the capacity of the beams. The CG error is less than 10 \% in terms of peak load (See the ninth column in Table \ref{tab:LD_comparison}) and, excluding beams B2 and B13, the agreement remains satisfactory even in the post-failure regime. 

Furthermore, the numerical results agree well, again with exception of beam B2, with the experimental data up to the peak load. In the post failure regime, however, the numerical response is less softening than the experimental one. This discrepancy is most likely due to the fact that the experiments show debonding of the reinforcement which is not included in the simulations.

For beam B2 the CG model approximates poorly the FS model because the size of the shear span is too small. Indeed, for B2 $a=200$ mm which is only 4 times the CG maximum aggregate size and the CG discretization does not have sufficient resolution for resolving adequately the shear crack formation. As far as the experiments are concerned, the reported stiffness of the beam is considerably lower than the ones simulated in the fine and coarse simulations, which however are accurate for the other beams. Also the B2 peak load is significantly lower than the corresponding capacity computed according to ACI specifications \citep{aci2014building}. This suggests that, for unknown reasons to the authors, the experimental curve for B2 is not fully reliable.

In agreement with the experimental evidence, the simulations predict that the first cracks are initiated in the constant moment region and propagate vertically in all beams (Fig. \ref{fig:CO_DBEAM}). This leads to the stiffness reduction visible in the initial portion of the load versus displacement curves in Fig. \ref{fig:LD_DBEAM}. Subsequently, additional flexural cracks are generated in the vicinity of the support plates. These cracks become diagonal and grow towards the loading plates. As the loading process continues, additional smaller secondary inclined cracks initiate at the beam mid-depth. For the beams B8, B11 and B12 the crack propagation and crack width increase and they ultimately lead to the yielding of the longitudinal tension reinforcement as indicated by a relatively long plateau in the load versus displacement curve (see Figs. \ref{fig:LD_DBEAM}b, d, and e and the crack pattern in Figs. \ref{fig:CO_DBEAM}c \& d, g \& h, and i \& j. Ultimately, failure occurs with a localized crushing in the compression zone near the loading plates. On the contrary, beams B2, B10 and B13 fail with the localization of a major diagonal crack within the shear span as shown in Figs. \ref{fig:CO_DBEAM}a \& b, e \& f, and k \& l. 

Comparison of the crack patterns and the load versus displacement curves show that the CG models can capture the beams failure well and that, although the crack resolution decreases in the coarsening process, the effect on the load versus displacement curves is negligible.

\begin{table}[t!]
\small
\centering
\caption {Peak load, peak deflection, simulation time, number of processors for the FS and CG model; CG error and computational gain.} \label{tab:LD_comparison}
\begin{tabular}{|l|c|ccc|ccc|c|c|}
\hline
    & $P^{ex}_0$ &$P_0$ & $T$  & $N_p$ & $P^\mathcal{C}_0$ & $T^\mathcal{C}$  & $N^\mathcal{C}_p$ & $err^\mathcal{C}$ & $\frac{N_p T}{N^\mathcal{C}_p T^\mathcal{C}}$\\
    & [kN]  &[kN] & [h] & [-] & [kN]  &[h] & [-] & [\%] & [-]\\
\hline
B2  & 1550   & 2,441 & 96 & 16 & 2,884  &  13 &  8 &  18.15&   15\\
B8  & 1501  & 1,522 &  94 & 32 &  1,660 &   6 &  16 &   9.07&   30\\
B10 & 616   &   854 &  96 & 32 &    871 &  15 &  16 &   1.94&   13\\ 
B11 & 1025   &   990 &  66 & 32 &  1,021 & 15 &  8 &   3.13&   18\\
B12 & 1161   & 1,001 &  64 & 32 &  1,039 & 12 &  8 &   3.80&   22\\
B13 & 2985   & 2,971 &  168 & 64 & 2,931 & 14 & 16 & 1.35&   49\\
\hline    
\end{tabular}
\end{table}

In addition to the accuracy of the CG procedure, it is important to evaluate the gain in computational cost. Table \ref{tab:LD_comparison} shows the wall time and number of processors for the FS simulations ($T$ and $N_p$) and the CG simulations ($T^\mathcal{C}$ and $N_p^\mathcal{C}$) as well as the computational gain calculated as $N_pT/N^\mathcal{C}_pT^\mathcal{C}$. It is worth noting that for these simulations as well as all the other simulations discussed in this paper the FS models ran on a MPI --Message Passing Interface, \citep{gropp1999using}-- supercomputing cluster whereas the CG models ran on the same system but with OMP --OpenMP, Open Multi-Processing \citep{chapman2008using}-- directives. 

As one can see the computational gain ranges from 13 to 30 for a coarsening factor of 2.5 and is equal to 49 for a coarsening factor of 5. This demonstrates the effectiveness of the CG procedure in reducing the computational cost while retaining satisfactory accuracy.

\subsection{Reinforced Concrete Column}
\begin{figure}[b!]
\centering
\includegraphics[width=0.9\linewidth]{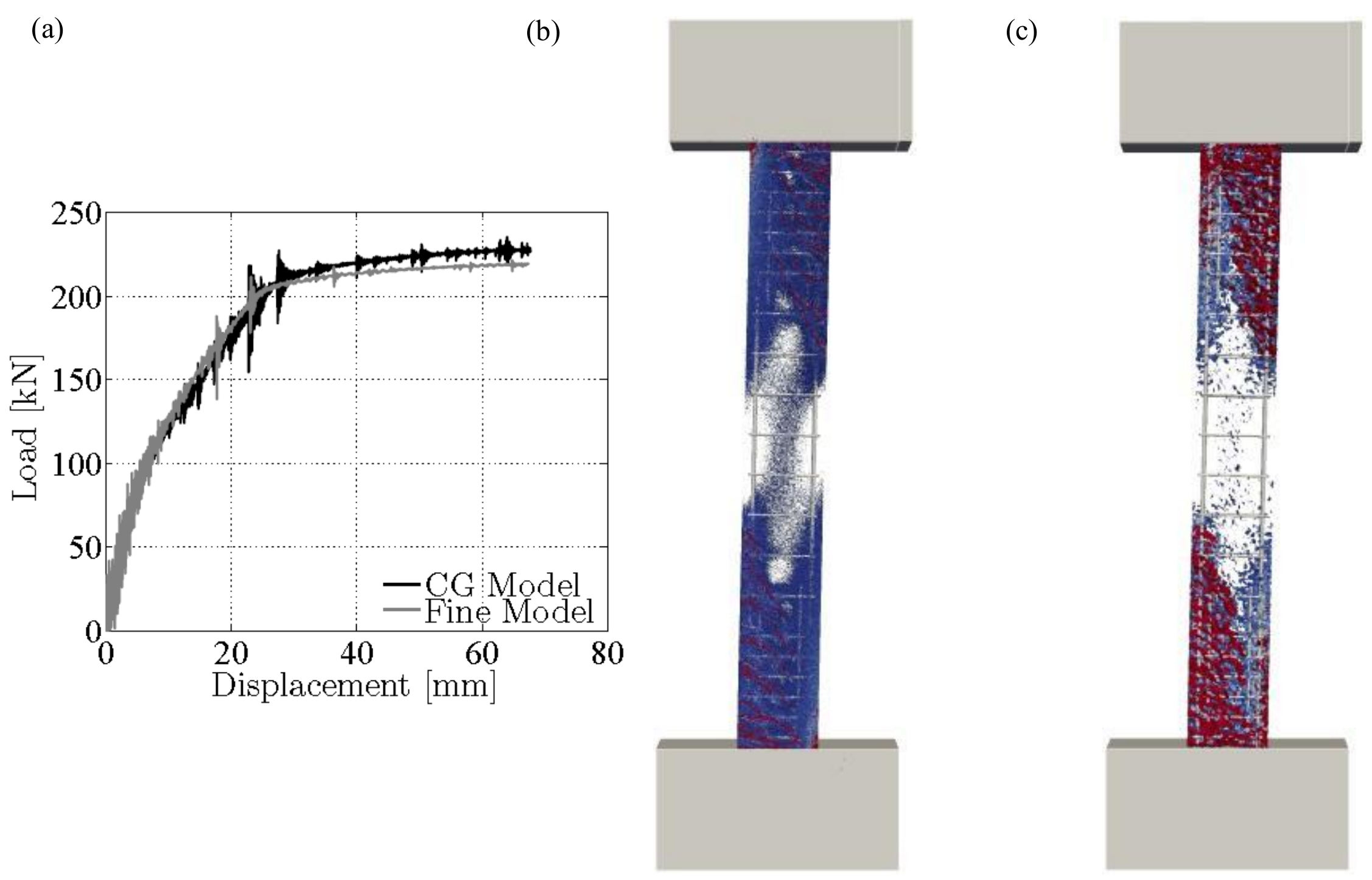}
\caption{(a) Load displacement curves for the reinforced concrete column. Crack pattern of the reinforced concrete column for the (b) FS model and (c) CG model.}
\label{fig:CO_Column}
\end{figure}

In this example, a concrete column subjected to lateral deflection is simulated with the FS and CG models for the C40 concrete. It is worth mentioning that in this case, the objective is not the comparison with the experiments but rather the assessment of accuracy and computational gain of the CG approach.

Dimensions and reinforcement details of the column are selected such that its behavior is totally controlled by bending. The concrete column is 3 m high with $300$ mm $\times$ $400$ mm  cross section. A reinforced concrete base $1,200$ mm wide, $300$ mm thick, and $600$ mm high is built integrally with the column to provide fixed boundary conditions at the bottom and at the top. The column longitudinal reinforcement consists of 8 $\phi 20_C$ rebars along the full length of the column and 2 $\phi 20_C$ additional rebars at each face of the column extended for 80 cm from both ends. $\phi 10_C$ closed hoop ties provide transverse reinforcement with 10 cm spacing along the first 50 cm length at the top and bottom of the column. The tie spacing is 20 cm along the rest of the column height. $\phi 10_C$ and $\phi 20_C$ are rebar types according to the Canadian standard and are characterized by nominal cross sectional areas equal to 100 mm$^2$ and 300 mm$^2$, respectively.

The column is first loaded  with a $600$ kN axial load and it is then loaded in the lateral direction with an applied displacement at the beam top end. The vertical load is kept constant during the lateral deflection. The top block rotations and the top block displacement orthogonal to the applied displacement are set to zero. The bottom block is completely fixed. The reinforcement is simulated in the same way as discussed in Section \ref{sec:deep-beams} with the following parameters: $E_s=200$GPa, $E_{sh}=1,372$MPa, $f_y=447$ GPa for the rebars; and $E_s=200$GPa, $E_{sh}=1,195$GPa, $f_y=455$ GPa for the ties.

The load versus displacement curves obtained in the FS and CG simulations match well as one can see  in Fig. \ref{fig:CO_Column}a. The agreement is excellent in all phases of the loading process: elastic regime, cracked regime with change in column stiffness, and final yielding of the rebars. For a lateral displacement of 67 mm the predicted FS load is 219 kN and the predicted CG load is 229 kN, which makes the CG error to be 4.6 \%. The predicted crack pattern of the FS and CG models at the end of the simulations are plotted in Fig. \ref{fig:CO_Column}b and c, respectively. One can see that the CG model reproduces qualitatively the cracking pattern observed in the FS simulation. However, the crack openings for the CG model are larger than the ones in the FS model. This is consistent with the coarse graining procedure since each CG crack represents the effects of multiple FS cracks. 

The FS model is characterized by 3,227,604 degrees of freedom and the simulation wall time was 150 hours with 64 MPI processors. On the contrary, the CG model had 45,312 degrees of freedom and the simulation wall time was 18.6 hours with 16 OMP processors. This leads to a  computational gain of 32.3.

\subsection{Reinforced Concrete Beam-Column Joint}
\begin{figure}[b!]
\centering
\includegraphics[trim={1cm 0 2cm 0},clip, width=1.0\linewidth]{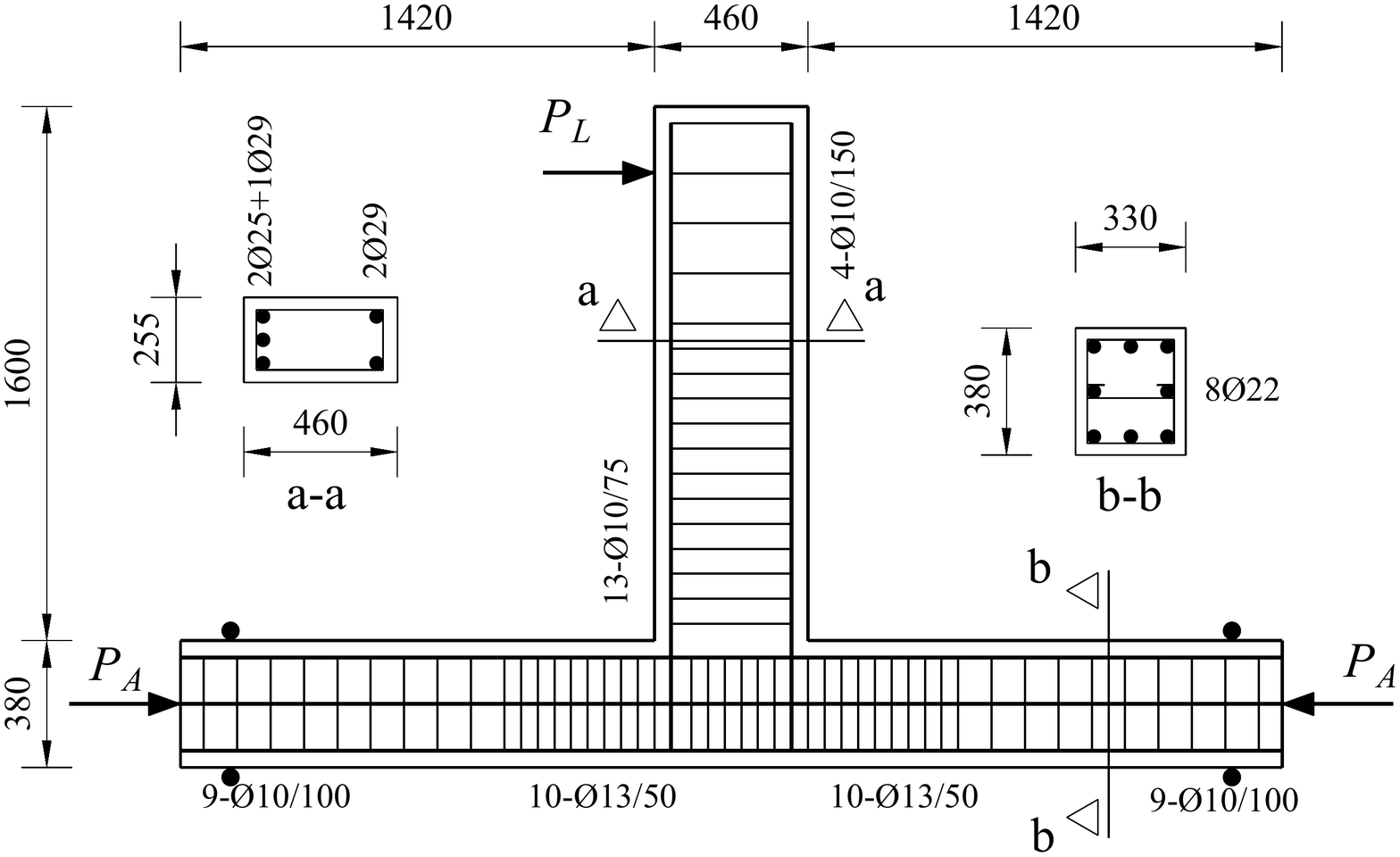}
\caption{Dimensions and reinforcement details of the exterior joints. All dimensions are in mm.}
\label{fig:GEOM_JOINTS}
\end{figure}
\begin{figure}[t]
\centering
\includegraphics[width=\linewidth]{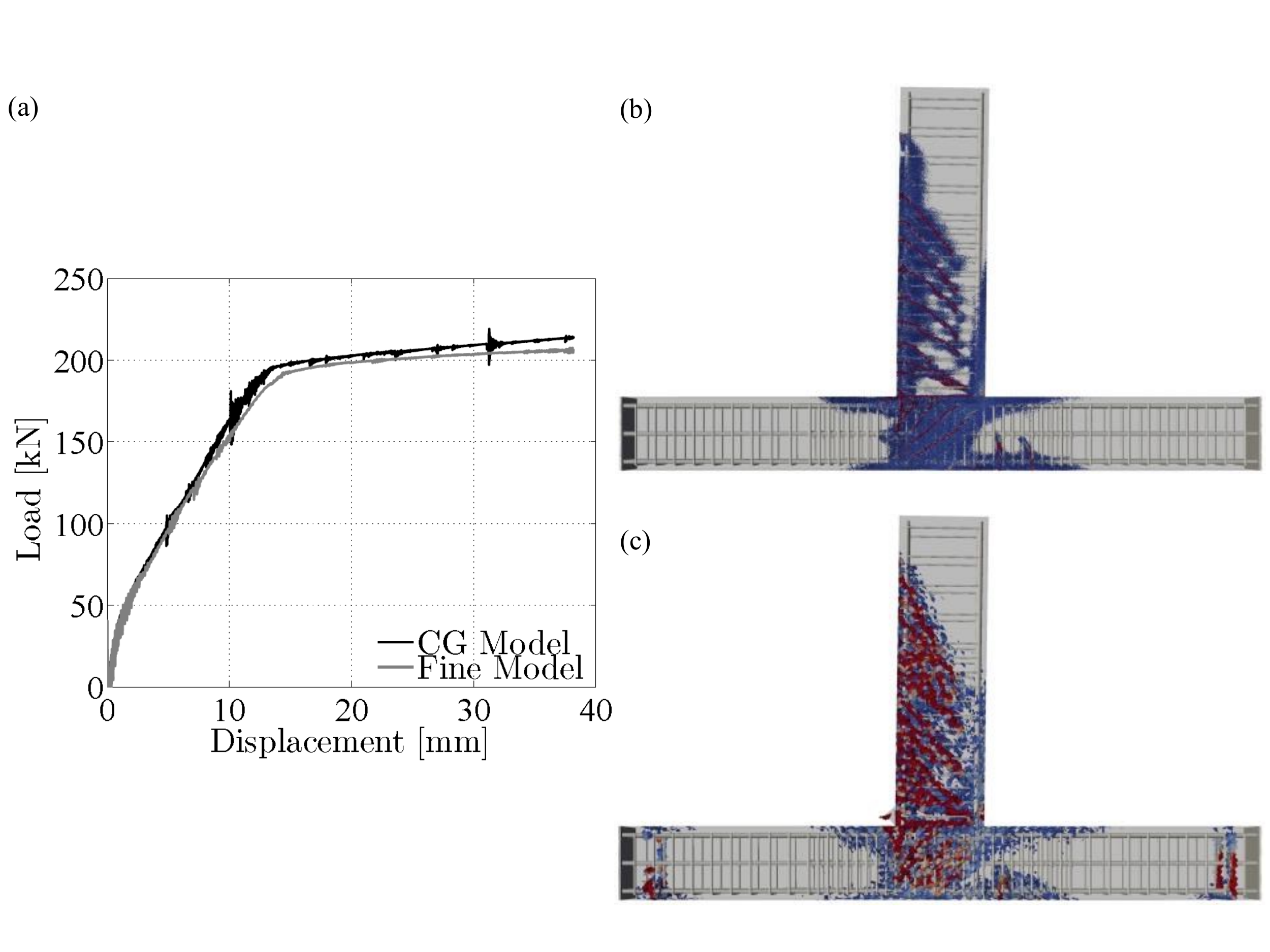}
\caption{(a) Load displacement curve for the exterior joinst. Crack pattern for exterior joint for (b) FS model and (c) CG model.}
\label{fig:CO_JOINTS}
\end{figure}
The full-scale exterior reinforced concrete beam-column joint tested by Megget \cite{megget1974cyclic}, was modeled in the current study by using concrete C40. Similarly to the previous example, here the goal is not the comparison with the experimental data but rather the assessment of accuracy and computational gain of the coarse graining approach.

This specimen represents realistically a typical exterior beam-column joint of a frame in an actual reinforced concrete building.

Dimensions of the beam-column joint and reinforcement details are given in Fig. \ref{fig:GEOM_JOINTS}. Similarly to the experiments, in the figure the specimen is placed such that the column is horizontal and the beam extends in the vertical direction. The column is supported  by rollers, which were simulated with linear elastic finite elements ($E=200$ GPa and $\nu=0.3$). The rollers were connected to the column via a penalty frictional contact algorithm governed by a friction coefficient equal to $\mu=0.3$ \citep{mars-1}.

The column was preloaded with a $200$ kN axial load and then the beam was loaded in the transverse direction under displacement control.

The flexural reinforcement consisted of 8 $\phi 22_A$ rebars for the column and 2 $\phi 25_A$ + 1 $\phi 29_A$ rebars for the beam at the top side, and 2 $\phi 29_A$ rebars for the beam at the bottom side. Furthermore, the column was reinforced with $\phi 10_A$ closed hoops with 100 mm spacing at the bottom and top thirds of column length. The central part of the column, on the contrary, had $\phi 13_A$ hoops with 50 mm spacing at the column confinement zones. The labels $\phi 10_A$, $\phi 13_A$, $\phi 22_A$, $\phi 25_A$, and $\phi 29_A$ are rebar types according to the American standard and they are characterized by the following cross sectional areas, yield stresses and hardening modulus:  71, 129, 387, 509, 645 mm$^2$; 409, 323, 372, 385, 382 MPa; 406, 400, 265, 1047, 708 respectively. For all rebars the Young's modulus $E_s=200$ MPa was used.
The beam had transverse reinforcement in the form of $\phi 10_A$ stirrups evenly spaced every 75 mm close to the column and every 150 mm at the beam loaded end. 

Figure \ref{fig:CO_JOINTS}a shows the obtained load versus beam tip displacement curves. One can see that the response of the CG simulation corresponds very well with the FS one in all phases of the loading process. For a displacement of 38 mm the FS and CG loads are 206 and 214 kN. This corresponds to a CG error of 3.9 \%. Figure \ref{fig:CO_JOINTS}b and c illustrate the crack pattern obtained in the FS and CG simulations, respectively: the CG model is again able to predict the damaged zones with high accuracy. 

The FS model had 5,382,372 degrees of freedom and ran 168 hours on 64 MPI processors; the CG model had 60,390 degrees of freedom and ran for 18.5 hours on 16 OMP processors. Consistently, the computational gain was 36.32.

\subsection{Reinforced Concrete Frame}
\label{sec:frame}
This section analyzes a single-span, two-story, shear-critical reinforced concrete frame tested at University of Toronto \cite{duong2006seismic}.
%

\begin{figure}[h!]
\centering
\includegraphics[width=1\linewidth]{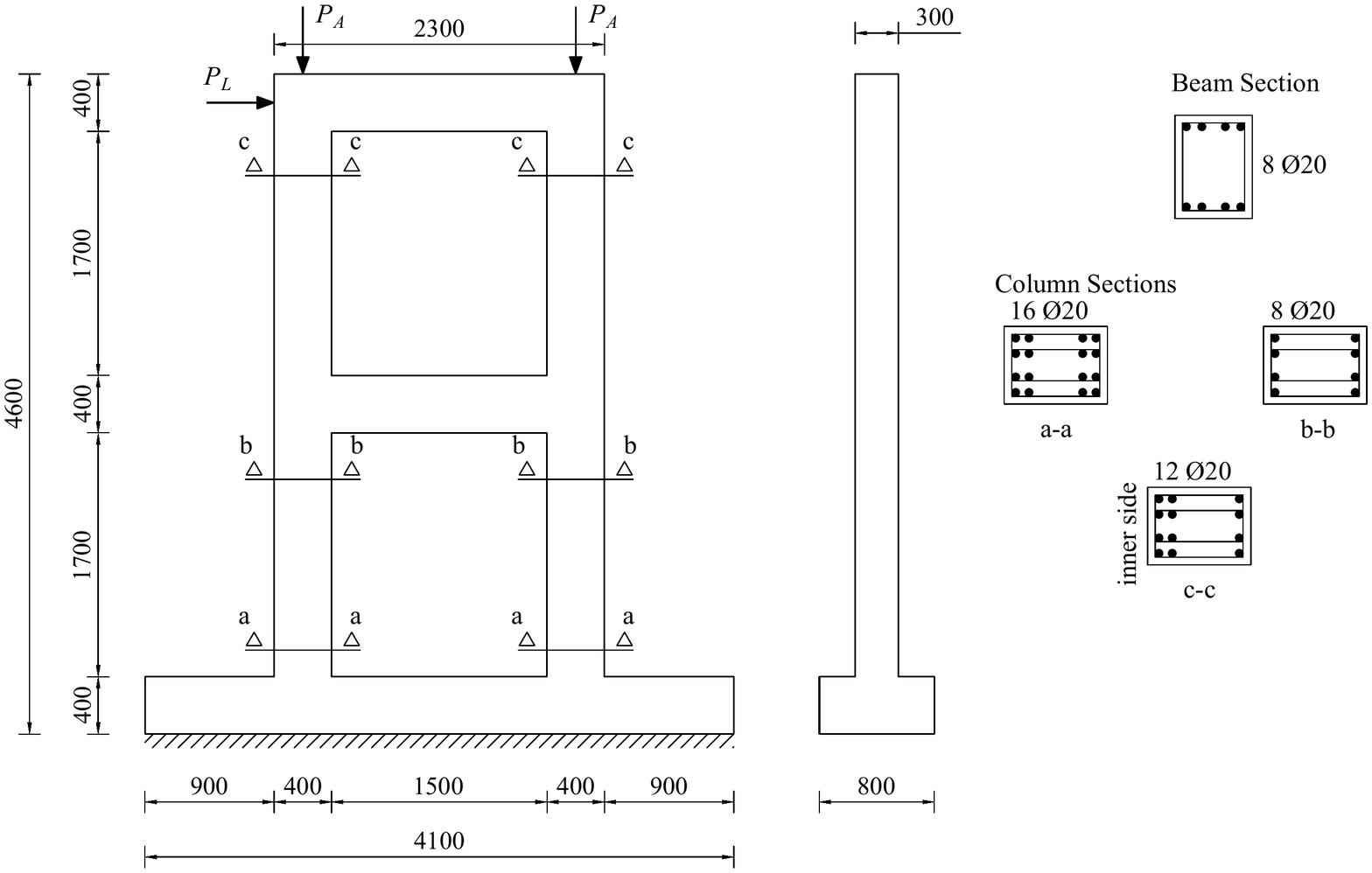}
\caption{Dimensions and reinforcement details of the concrete frame. All values are reported in mm.}
\label{fig:GEOM_FRAME}
\end{figure}

Figure \ref{fig:GEOM_FRAME} shows all geometrical dimensions of the frame and reinforcement details. Beams and columns have all the same 300 $\times$ 400 mm cross section. The span-to-depth ratio of the beams is $3.75$, which was selected to provide shear critical behavior. A reinforced concrete base of $800$ mm width, $400$ mm thickness, and $4100$ mm long was integrated with the frame to provide fixed boundary conditions at the bottom. Three different types of reinforcing rebars were used in the frame, namely $\phi 20_C$ for the flexural reinforcement; $\phi 10_C$ for shear reinforcement in the columns; and $\phi 10_A$ for shear reinforcement in the beams. 
The properties of these rebars were introduced earlier in this paper. Single stirrups were used in the beams and they were spaced evenly every $300$ mm; double stirrups were used in the columns and they were spaced evenly every $130$ mm in the columns (see Fig. \ref{fig:GEOM_FRAME}-right). The concrete base included 8 $\phi 20_A$ top and bottom bars with $\phi$10$_c$ triple closed stirrups spaced uniformly at $175$ mm.

As illustrated in Fig. \ref{fig:GEOM_FRAME}, the frame was preloaded in the vertical direction with a $420$ kN force applied to each column by using $300 \times 300 \times 25$ mm loading plates. The vertical loading was held constant throughout the test. The frame was then loaded in the horizontal direction at the level of the top story beam by using a displacement controlled actuator and a $254 \times 254 \times 25$ mm steel plate.

\begin{figure}[t!]
\centering 
{\includegraphics[width=0.9\textwidth]{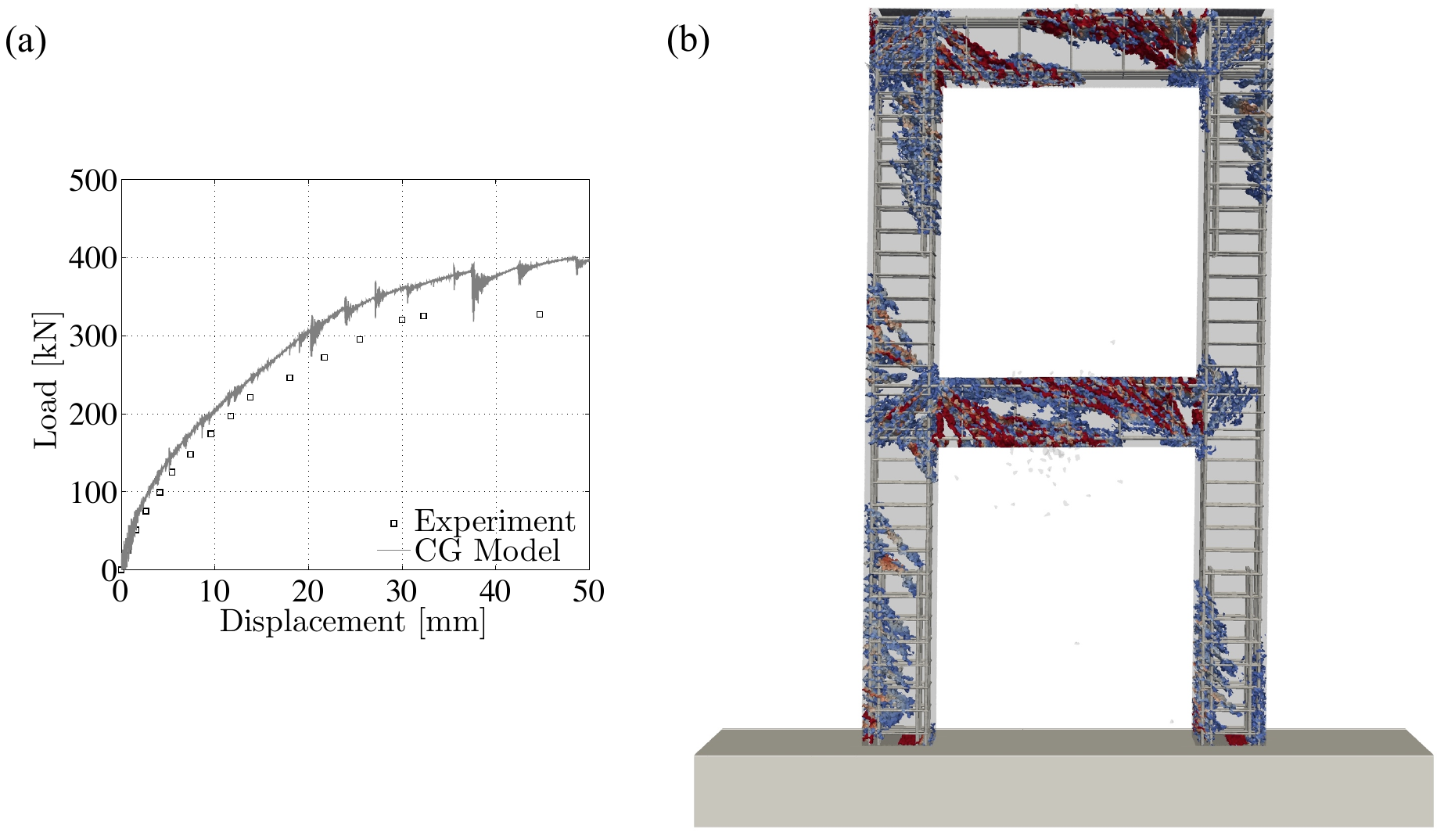}}
\caption{(a) Load displacement curve. (b) Crack pattern of the RC frame.}
\label{fig:FRAME_FIG}
\end{figure}

In order to reduce the computational cost, the concrete base was modeled by elastic hexahedral finite elements ($E=40$ GPa and $\nu=0.2$) and only the frame was modeled by LDPM. The reinforcement was simulated in the same way as in the previously discussed examples.  

The loading plates were modeled with elastic quadratic shell elements ($E=200$ GPa and $\nu=0.3$) and were connected to LDPM nodes using penalty constraints \citep{mars-1}. The displacements at the bottom face of the concrete base were fixed in all directions. The numerical simulations were carried out only with the C40 CG model (194,946 LDPM degrees of freedom) because the FS model consisted of more than 10 million degrees of freedom and running such a large model would have required an enormous computational cost. Therefore, in this example the results are compared to the experimental data.

The load versus displacement at the top joint obtained from the coarse grained simulation is plotted in Fig. \ref{fig:FRAME_FIG}a along with the experimental results. The maximum lateral load reached during the experiment is reported as 327 kN, which compares well with 395 kN obtained from the CG simulation at same drift ratio. The CG simulation ran in 48 hours by using 32 MPI processors.

The simulation overestimates the load capacity by approximately 20\%, which is still acceptable considering the typical scatter or experimental data (only one test was performed) and the significant gain of the computational cost. Also, the experiments reported evidence of rebar debonding at the base of the left-hand-side column whereas, as mentioned earlier in this paper, such debonding was not included in the simulations


Finally, the failure mode of the frame is captured well in the LDPM simulation (Fig. \ref{fig:FRAME_FIG}b), in which it is evident that shear cracking and failure of the beams govern the frame behavior as reported in the experiments.

\section{Conclusions}
In this paper a homogenization based coarse graining technique for the multiscale analysis of reinforced concrete structures was formulated within the framework of the Lattice Discrete Particle Model. 

This was achieved by approximating the RVE homogenized response of the fine scale LDPM with a coarse grained LDPM. The approximation was performed by optimizing the material parameters of the coarse grained model using an automatic parameter identification technique based on the nonlinear least square method. The proposed model was validated by means of several numerical simulations on various structural systems including: reinforced concrete deep beams of various sizes, a concrete column, a concrete beam-column joints, and a reinforced concrete frame. 

On the basis of the simulations discussed in this paper, the following conclusions can be drawn.

\begin{enumerate}
\item The computational gain expected with the formulated coarse grained approach depends on the size of the coarsening factor. Indeed the larger the coarsening factor is, the smaller is the number of degrees of freedom in the coarse grained system. Hence, the coarse grained model requires fewer operations for each loading step.
\item If the equations motion of the LDPM cells are solved with an explicit scheme, additional computational saving arises from the fact that the coarser discretization allows for a bigger stable time step reducing the total number of steps required for a certain loading history.
\item The size of the coarsening factor is limited by the size of the structural element to be simulated and by the loading condition. The coarse grained system must have enough resolution to simulate accurately the main features of structural system response.
\item To simulate correctly softening behavior, the size of the fine scale RVE and the coarse grained RVE must have the same size. This requirement, along with the fact that the RVE must be at least 5 times the maximum particle size, limits the coarsening factor because a critical size of the fine scale RVE exists at which the fine scale response features snap back and it cannot be used for the identification of the coarse grained LDPM parameters.
\item Because of the limitations discussed above the maximum coarsening factor is about 5 for standard concrete mixes. Although not pursued in this paper, a larger coarsening factor can be achieved if the coarse grained parameters are identified sequentially by using a series of intermediate coarsening factors. 
\item For a coarsening factor equal to 2.5 and 5 and for fine scale LDPM parameters relevant to standard concrete mixes, the computational gain is in the order of 20 and 50, respectively, with an error on the predicted ultimate loads smaller than 10 \%.
\end{enumerate}



\section*{Acknowledgments}
Financial support from the U.S. National Science Foundation (NSF) under Grant No. CMMI-1435923 is gratefully acknowledged. The work of the first author was also partially supported
by the Scientific and Technological Research Council of Turkey (TUBITAK).





\begin{thebibliography}{10}
	\expandafter\ifx\csname url\endcsname\relax
	\def\url#1{\texttt{#1}}\fi
	\expandafter\ifx\csname urlprefix\endcsname\relax\def\urlprefix{URL }\fi
	\expandafter\ifx\csname href\endcsname\relax
	\def\href#1#2{#2} \def\path#1{#1}\fi
	
	\bibitem{roelfstra1985beton}
	P.~Roelfstra, H.~Sadouki, F.~Wittmann, Le b{\'e}ton num{\'e}rique, Materials
	and Structures 18~(5) (1985) 327--335.
	
	\bibitem{caballero2006meso}
	A.~Caballero, I.~Carol, C.~L{\'o}pez, A meso-level approach to the 3d numerical
	analysis of cracking and fracture of concrete materials, Fatigue \& Fracture
	of Engineering Materials \& Structures 29~(12) (2006) 979--991.
	
	\bibitem{caballero2006new}
	A.~Caballero, I.~Carol, C.~Lopez, New results in 3d meso-mechanical analysis of
	concrete specimens using interface elements, in: Computational modelling of
	concrete structures, EURO-C, 2006, pp. 43--52.
	
	\bibitem{caballero20063d}
	A.~Caballero, C.~L{\'o}pez, I.~Carol, 3d meso-structural analysis of concrete
	specimens under uniaxial tension, Computer Methods in Applied Mechanics and
	Engineering 195~(52) (2006) 7182--7195.
	
	\bibitem{schlangen1992simple}
	E.~Schlangen, J.~Van~Mier, Simple lattice model for numerical simulation of
	fracture of concrete materials and structures, Materials and Structures
	25~(9) (1992) 534--542.
	
	\bibitem{bolander1998fracture}
	J.~Bolander~Jr, S.~Saito, Fracture analyses using spring networks with random
	geometry, Engineering Fracture Mechanics 61~(5-6) (1998) 569--591.
	
	\bibitem{yip2006irregular}
	M.~Yip, Z.~Li, B.-S. Liao, J.~Bolander, Irregular lattice models of fracture of
	multiphase particulate materials, International journal of fracture 140~(1-4)
	(2006) 113--124.
	
	\bibitem{nagai2004mesoscopic}
	K.~Nagai, Y.~Sato, T.~Ueda, Mesoscopic simulation of failure of mortar and
	concrete by 2d rbsm, Journal of Advanced Concrete Technology 2~(3) (2004)
	359--374.
	
	\bibitem{cundall1979discrete}
	P.~A. Cundall, O.~D. Strack, A discrete numerical model for granular
	assemblies, geotechnique 29~(1) (1979) 47--65.
	
	\bibitem{plesha1983modeling}
	M.~Plesha, E.~Aifatis, et~al., On the modeling of rocks with microstructure,
	in: The 24th US Symposium on Rock Mechanics (USRMS), American Rock Mechanics
	Association, 1983.
	
	\bibitem{zubelewicz1987interface}
	A.~Zubelewicz, Z.~P. Ba{\v{z}}ant, Interface element modeling of fracture in
	aggregate composites, Journal of engineering mechanics 113~(11) (1987)
	1619--1630.
	
	\bibitem{cusatis2003confinement}
	G.~Cusatis, Z.~P. Ba{\v{z}}ant, L.~Cedolin, Confinement-shear lattice model for
	concrete damage in tension and compression: I. theory, Journal of Engineering
	Mechanics 129~(12) (2003) 1439--1448.
	
	\bibitem{Donze-discrete}
	N.~Belheine, J.~Plassiard, F.~Donz\'e, F.~Darve, A.~Seridi, Numerical
	simulation of drained triaxial test using 3d discrete element modeling,
	Comput Geotech 36~(1-2) (2009) 320--331.
	
	\bibitem{cusatis2003confinementII}
	G.~Cusatis, Z.~P. Ba{\v{z}}ant, L.~Cedolin, Confinement-shear lattice model for
	concrete damage in tension and compression. ii. computation and validation,
	Journal of Engineering Mechanics 129~(12) (2003) 1449--1458.
	
	\bibitem{cusatis-ldpm-1}
	G.~Cusatis, D.~Pelessone, A.~Mencarelli, Lattice discrete particle model (ldpm)
	for failure behavior of concrete. i: Theory, Cement and Concrete Composites
	33~(9) (2011) 881--890.
	
	\bibitem{cusatis-ldpm-2}
	G.~Cusatis, A.~Mencarelli, D.~Pelessone, J.~Baylot, Lattice discrete particle
	model (ldpm) for failure behavior of concrete. ii: Calibration and
	validation, Cement and Concrete composites 33~(9) (2011) 891--905.
	
	\bibitem{Alnaggar2012automatic}
	M.~Alnaggar, G.~Cusatis, Automatic parameter identification of discrete
	mesoscale models with application to the coarse-grained simulation of
	reinforced concrete structures, in: 20th Analysis and computation specialty
	conference, Vol.~36, 2012, pp. 406--417.
	
	\bibitem{gitman2007representative}
	I.~Gitman, H.~Askes, L.~Sluys, Representative volume: existence and size
	determination, Engineering fracture mechanics 74~(16) (2007) 2518--2534.
	
	\bibitem{kouznetsova2004size}
	V.~Kouznetsova, M.~Geers, W.~Brekelmans, Size of a representative volume
	element in a second-order computational homogenization framework,
	International Journal for Multiscale Computational Engineering 2~(4).
	
	\bibitem{Bostanabad-1}
	R.~Bostanabad, Y.~Zhang, X.~Li, T.~Kearney, L.~Brinson, D.~Apley, W.~Liu,
	W.~Chen, Computational microstructure characterization and reconstruction:
	Review of the state-of-the-art techniques, Progress in Materials Science 95
	(2018) 1--41.
	
	\bibitem{Bessa-1}
	M.~Bessa, R.~Bostanabad, Z.~Liu, A.~Hu, D.~Apley, C.~Brinson, W.~Chen, W.~Liu,
	A framework for data-driven analysis of materials under uncertainty:
	Countering the curse of dimensionality, Computer Methods in Applied Mechanics
	and Engineering 320 (2017) 633--667.
	
	\bibitem{smit1998prediction}
	R.~Smit, W.~Brekelmans, H.~Meijer, Prediction of the mechanical behavior of
	nonlinear heterogeneous systems by multi-level finite element modeling,
	Computer methods in applied mechanics and engineering 155~(1-2) (1998)
	181--192.
	
	\bibitem{miehe1999computational}
	C.~Miehe, J.~Schr{\"o}der, J.~Schotte, Computational homogenization analysis in
	finite plasticity simulation of texture development in polycrystalline
	materials, Computer methods in applied mechanics and engineering 171~(3-4)
	(1999) 387--418.
	
	\bibitem{Hasani-1}
	B.~Hassani, E.~Hinton, A review of homogenization and topology optimization i:
	homogenization theory for media with periodic structure, Computers \&
	Structures 69 (1998) 707--717.
	
	\bibitem{Hasani-2}
	B.~Hassani, E.~Hinton, A review of homogenization and topology optimization ii:
	analytical and numerical solution of homogenization equations, Computers \&
	Structures 69 (1998) 719--738.
	
	\bibitem{fish2007generalized}
	J.~Fish, W.~Chen, R.~Li, Generalized mathematical homogenization of atomistic
	media at finite temperatures in three dimensions, Computer methods in applied
	mechanics and engineering 196~(4) (2007) 908--922.
	
	\bibitem{rezakhani2016asymptotic}
	R.~Rezakhani, G.~Cusatis, Asymptotic expansion homogenization of discrete
	fine-scale models with rotational degrees of freedom for the simulation of
	quasi-brittle materials, Journal of the Mechanics and Physics of Solids 88
	(2016) 320--345.
	
	\bibitem{rezakhani2017adaptive}
	R.~Rezakhani, X.~Zhou, G.~Cusatis, Adaptive multiscale homogenization of the
	lattice discrete particle model for the analysis of damage and fracture in
	concrete, International Journal of Solids and Structures 125 (2017) 50--67.
	
	\bibitem{muller2002coarse}
	F.~M{\"u}ller-Plathe, Coarse-graining in polymer simulation: From the atomistic
	to the mesoscopic scale and back, ChemPhysChem 3~(9) (2002) 754--769.
	
	\bibitem{noid2008multiscale-I}
	W.~Noid, J.-W. Chu, G.~S. Ayton, V.~Krishna, S.~Izvekov, G.~A. Voth, A.~Das,
	H.~C. Andersen, The multiscale coarse-graining method. i. a rigorous bridge
	between atomistic and coarse-grained models, The Journal of chemical physics
	128~(24) (2008) 244114.
	
	\bibitem{noid2008multiscale-II}
	W.~Noid, P.~Liu, Y.~Wang, J.-W. Chu, G.~S. Ayton, S.~Izvekov, H.~C. Andersen,
	G.~A. Voth, The multiscale coarse-graining method. ii. numerical
	implementation for coarse-grained molecular models, The Journal of chemical
	physics 128~(24) (2008) 244115.
	
	\bibitem{cranford2010coarse}
	S.~Cranford, M.~J. Buehler, Coarse-graining parameterization and multiscale
	simulation of hierarchical systems. part i: Theory and model formulation,
	Tech. rep., DTIC Document (2010).
	
	\bibitem{PhysRevLett.105.237802}
	S.~O. Nielsen, P.~B. Moore, B.~Ensing, Adaptive multiscale molecular dynamics
	of macromolecular fluids, Phys. Rev. Lett. 105 (2010) 237802.
	\newblock \href {http://dx.doi.org/10.1103/PhysRevLett.105.237802}
	{\path{doi:10.1103/PhysRevLett.105.237802}}.
	
	\bibitem{praprotnik2005adaptive}
	M.~Praprotnik, L.~Delle~Site, K.~Kremer, Adaptive resolution molecular-dynamics
	simulation: Changing the degrees of freedom on the fly, The Journal of
	chemical physics 123~(22) (2005) 224106.
	
	\bibitem{rzepiela2011hybrid}
	A.~J. Rzepiela, M.~Louhivuori, C.~Peter, S.~J. Marrink, Hybrid simulations:
	combining atomistic and coarse-grained force fields using virtual sites,
	Physical Chemistry Chemical Physics 13~(22) (2011) 10437--10448.
	
	\bibitem{brini2013systematic}
	E.~Brini, E.~A. Algaer, P.~Ganguly, C.~Li, F.~Rodr{\'\i}guez-Ropero, N.~F.
	van~der Vegt, Systematic coarse-graining methods for soft matter
	simulations--a review, Soft Matter 9~(7) (2013) 2108--2119.
	
	\bibitem{henderson1974uniqueness}
	R.~Henderson, A uniqueness theorem for fluid pair correlation functions,
	Physics Letters A 49~(3) (1974) 197--198.
	
	\bibitem{reith2003deriving}
	D.~Reith, M.~P{\"u}tz, F.~M{\"u}ller-Plathe, Deriving effective mesoscale
	potentials from atomistic simulations, Journal of computational chemistry
	24~(13) (2003) 1624--1636.
	
	\bibitem{PhysRevE.52.3730}
	A.~P. Lyubartsev, A.~Laaksonen, Calculation of effective interaction potentials
	from radial distribution functions: A reverse monte carlo approach, Phys.
	Rev. E 52 (1995) 3730--3737.
	\newblock \href {http://dx.doi.org/10.1103/PhysRevE.52.3730}
	{\path{doi:10.1103/PhysRevE.52.3730}}.
	
	\bibitem{ercolessi1994interatomic}
	F.~Ercolessi, J.~B. Adams, Interatomic potentials from first-principles
	calculations: the force-matching method, EPL (Europhysics Letters) 26~(8)
	(1994) 583.
	
	\bibitem{izvekov2005multiscale}
	S.~Izvekov, G.~A. Voth, A multiscale coarse-graining method for biomolecular
	systems, The Journal of Physical Chemistry B 109~(7) (2005) 2469--2473.
	
	\bibitem{izvekov2005systematic}
	S.~Izvekov, A.~Violi, G.~A. Voth, Systematic coarse-graining of nanoparticle
	interactions in molecular dynamics simulation, The Journal of Physical
	Chemistry B 109~(36) (2005) 17019--17024.
	
	\bibitem{mullinax2009generalized}
	J.~Mullinax, W.~Noid, A generalized-yvon- born- green theory for determining
	coarse-grained interaction potentials†, The Journal of Physical Chemistry C
	114~(12) (2009) 5661--5674.
	
	\bibitem{cusatis2013high}
	G.~Cusatis, X.~Zhou, High-order microplane theory for quasi-brittle materials
	with multiple characteristic lengths, Journal of Engineering Mechanics
	140~(7) (2013) 04014046.
	
	\bibitem{lale2017isogeometric}
	E.~Lale, X.~Zhou, G.~Cusatis, Isogeometric implementation of high-order
	microplane model for the simulation of high-order elasticity, softening, and
	localization, Journal of Applied Mechanics 84~(1) (2017) 011005.
	
	\bibitem{cusatis2017discontinuous}
	G.~Cusatis, R.~Rezakhani, E.~A. Schauffert, Discontinuous cell method (dcm) for
	the simulation of cohesive fracture and fragmentation of continuous media,
	Engineering Fracture Mechanics 170 (2017) 1--22.
	
	\bibitem{cusatis2006confinement}
	G.~Cusatis, Z.~P. Ba{\v{z}}ant, L.~Cedolin, Confinement-shear lattice csl model
	for fracture propagation in concrete, Computer methods in applied mechanics
	and engineering 195~(52) (2006) 7154--7171.
	
	\bibitem{bavzant1983crack}
	Z.~P. Ba{\v{z}}ant, B.~H. Oh, Crack band theory for fracture of concrete,
	Mat{\'e}riaux et construction 16~(3) (1983) 155--177.
	
	\bibitem{ceccato2017simulation}
	C.~Ceccato, M.~Salviato, C.~Pellegrino, G.~Cusatis, Simulation of concrete
	failure and fiber reinforced polymer fracture in confined columns with
	different cross sectional shape, International Journal of Solids and
	Structures 108 (2017) 216--229.
	
	\bibitem{cusatis-Jovanca}
	J.~Smith, G.~Cusatis, D.~Pelessone, E.~Landis, J.~O'Daniel, J.~Baylot, Discrete
	modeling of ultra-high-performance concrete with application to projectile
	penetration, International Journal of Impact Engineering 65 (2014) 13--32.
	
	\bibitem{mars-1}
	D.~Pelessone, Mars: Modeling and analysis of the response of
	structures-user’s manual, ES3, Beach (CA), USA.
	
	\bibitem{smith2017numerical}
	J.~Smith, G.~Cusatis, Numerical analysis of projectile penetration and
	perforation of plain and fiber reinforced concrete slabs, International
	Journal for Numerical and Analytical Methods in Geomechanics 41~(3) (2017)
	315--337.
	
	\bibitem{cusatis-mohammed}
	M.~Alnaggar, G.~Cusatis, G.~Di~Luzio, Lattice discrete particle modeling (ldpm)
	of alkali silica reaction (asr) deterioration of concrete structures, Cement
	and Concrete Composites 41 (2013) 45--59.
	
	\bibitem{alnaggar2017modeling}
	M.~Alnaggar, G.~Di~Luzio, G.~Cusatis, Modeling time-dependent behavior of
	concrete affected by alkali silica reaction in variable environmental
	conditions, Materials 10~(5) (2017) 471.
	
	\bibitem{pathirage2018effect}
	M.~Pathirage, F.~Bousikhane, M.~D’Ambrosia, M.~Alnaggar, G.~Cusatis, Effect
	of alkali silica reaction on the mechanical properties of aging mortar bars:
	Experiments and numerical modeling, International Journal of Damage Mechanics
	(2018) 1056789517750213.
	
	\bibitem{cusatis-Ed1}
	E.~A. Schauffert, G.~Cusatis, Lattice discrete particle model for
	fiber-reinforced concrete. i: theory, Journal of Engineering Mechanics
	138~(7) (2011) 826--833.
	
	\bibitem{cusatis-Ed2}
	E.~A. Schauffert, G.~Cusatis, D.~Pelessone, J.~L. O'Daniel, J.~T. Baylot,
	Lattice discrete particle model for fiber-reinforced concrete. ii: tensile
	fracture and multiaxial loading behavior, Journal of Engineering Mechanics
	138~(7) (2011) 834--841.
	
	\bibitem{jin2016lattice}
	C.~Jin, N.~Buratti, M.~Stacchini, M.~Savoia, G.~Cusatis, Lattice discrete
	particle modeling of fiber reinforced concrete: Experiments and simulations,
	European Journal of Mechanics-A/Solids 57 (2016) 85--107.
	
	\bibitem{wan2016analysis}
	L.~Wan, R.~Wendner, B.~Liang, G.~Cusatis, Analysis of the behavior of ultra
	high performance concrete at early age, Cement and Concrete Composites 74
	(2016) 120--135.
	
	\bibitem{wan2018age}
	L.~Wan-Wendner, R.~Wan-Wendner, G.~Cusatis, Age-dependent size effect and
	fracture characteristics of ultra-high performance concrete, Cement and
	Concrete Composites 85 (2018) 67--82.
	
	\bibitem{wan2016novel}
	L.~Wan, R.~Wendner, G.~Cusatis, A novel material for in situ construction on
	mars: experiments and numerical simulations, Construction and Building
	Materials 120 (2016) 222--231.
	
	\bibitem{ashari2017lattice}
	S.~E. Ashari, G.~Buscarnera, G.~Cusatis, A lattice discrete particle model for
	pressure-dependent inelasticity in granular rocks, International Journal of
	Rock Mechanics and Mining Sciences 91 (2017) 49--58.
	
	\bibitem{li2017multiscale}
	W.~Li, R.~Rezakhani, C.~Jin, X.~Zhou, G.~Cusatis, A multiscale framework for
	the simulation of the anisotropic mechanical behavior of shale, International
	Journal for Numerical and Analytical Methods in Geomechanics 41~(14) (2017)
	1494--1522.
	
	\bibitem{ceccato2018proper}
	C.~Ceccato, X.~Zhou, D.~Pelessone, G.~Cusatis, Proper orthogonal decomposition
	framework for the explicit solution of discrete systems with softening
	response, Journal of Applied Mechanics.
	
	\bibitem{Dennis}
	J.~Dennis, D.~Jacobs, State of the art in numerical analysis, State of the Art
	in Numerical Analysis.
	
	\bibitem{Bolzon}
	G.~Bolzon, R.~Fedele, G.~Maier, Parameter identification of a cohesive crack
	model by kalman filter, Computer Methods in Applied Mechanics and Engineering
	191~(25) (2002) 2847--2871.
	
	\bibitem{Aguir}
	H.~Aguir, H.~BelHadjSalah, R.~Hambli, Parameter identification of an
	elasto-plastic behaviour using artificial neural networks--genetic algorithm
	method, Materials \& Design 32~(1) (2011) 48--53.
	
	\bibitem{bazant1997fracture}
	Z.~P. Bazant, J.~Planas, Fracture and size effect in concrete and other
	quasibrittle materials, Vol.~16, CRC press, 1997.
	
	\bibitem{Smith201413}
	J.~Smith, G.~Cusatis, D.~Pelessone, E.~Landis, J.~O'Daniel, J.~Baylot, Discrete
	modeling of ultra-high-performance concrete with application to projectile
	penetration, International Journal of Impact Engineering 65~(0) (2014) 13 --
	32.
	\newblock \href
	{http://dx.doi.org/http://dx.doi.org/10.1016/j.ijimpeng.2013.10.008}
	{\path{doi:http://dx.doi.org/10.1016/j.ijimpeng.2013.10.008}}.
	
	\bibitem{kosasize}
	K.~Kosa, S.~Uchida, T.~Nishioka, H.~Kobayashi, Size effect on the shear strengh
	of rc deep beams.
	
	\bibitem{duong2006seismic}
	K.~V. Duong, Seismic behaviour of a shear-critical reinforced concrete frame:
	an experimental and numerical investigation, Ph.D. thesis, University of
	Toronto, Canada (2006).
	
	\bibitem{aci2014building}
	A.~C. 318, Building code requirements for structural concrete (aci 318-14) and
	commentary, American Concrete Institute, 2014.
	
	\bibitem{ismail2016shear}
	K.~S. Ismail, Shear behaviour of reinforced concrete deep beams, Ph.D. thesis,
	University of Sheffield, England (2016).
	
	\bibitem{bazant1984size}
	Z.~P. Bazant, J.~K. Kim, Size effect in shear failure of longitudinally
	reinforced beams., Journal of the American Concrete Institute 81~(5) (1984)
	456--468.
	
	\bibitem{bazant1987size}
	Z.~P. Bazant, H.-H. Sun, Size effect in diagonal shear failure: Influence of
	aggregate size and stirrups, ACI Materials Journal 84~(4) (1987) 259--272.
	
	\bibitem{bazant1991size}
	Z.~P. Bazant, M.~T. Kazemi, Size effect on diagonal shear failure of beams
	without stirrups, ACI Structural journal 88~(3) (1991) 268--276.
	
	\bibitem{frosch2017unified}
	R.~J. Frosch, Q.~Yu, G.~Cusatis, Z.~P. Ba{\v{z}}ant, A unified approach to
	shear design, Concrete International 39~(9) (2017) 47--52.
	
	\bibitem{yu2016comparison}
	Q.~Yu, J.-L. Le, M.~H. Hubler, R.~Wendner, G.~Cusatis, Z.~P. Ba{\v{z}}ant,
	Comparison of main models for size effect on shear strength of reinforced and
	prestressed concrete beams, Structural Concrete 17~(5) (2016) 778--789.
	
	\bibitem{tan1999shear}
	K.~Tan, H.~Lu, Shear behavior of large reinforced concrete deep beams and code
	comparisons, Structural Journal 96~(5) (1999) 836--846.
	
	\bibitem{jirasek2002inelastic}
	M.~Jir{\'a}sek, Z.~P. Bazant, Inelastic analysis of structures, John Wiley \&
	Sons, 2002.
	
	\bibitem{Alnaggar_PHD}
	M.~Alnaggar, Multiscale modeling of aging and deterioration of reinforced
	concrete structures, Ph.D. thesis, NORTHWESTERN UNIVERSITY, Evanston, IL, USA
	(2014).
	
	\bibitem{gropp1999using}
	W.~Gropp, E.~Lusk, A.~Skjellum, Using MPI: portable parallel programming with
	the message-passing interface, Vol.~1, MIT press, 1999.
	
	\bibitem{chapman2008using}
	B.~Chapman, G.~Jost, R.~Van Der~Pas, Using OpenMP: portable shared memory
	parallel programming, Vol.~10, MIT press, 2008.
	
	\bibitem{megget1974cyclic}
	L.~Megget, Cyclic behaviour of exterior reinforced concrete beam--column
	joints, Bulletin of the New Zealand National Society for Earthquake
	Engineering 7~(1) (1974) 27--47.
	
\end{thebibliography}

\end{document}